\documentclass[]{elsarticle}
\usepackage[utf8]{inputenc}
\usepackage[T1]{fontenc}
\usepackage{graphicx}
\usepackage{comment}
\usepackage{natbib}
\usepackage{amsmath}
\usepackage{amssymb}
\usepackage{lineno, comment, url}
\usepackage[usenames,dvipsnames,svgnames,table]{xcolor}
\usepackage{ulem}
\usepackage{lmodern}
\usepackage{todonotes}
\usepackage{setspace}
\usepackage[text={17.5cm,22cm},centering]{geometry}

\usepackage[english]{babel}

\newcommand{\rhob}{\text{\sout{\ensuremath{\rho}}}}
\newcommand{\uu}{\mathbf{u}}
\newcommand{\mm}{\mathbf{m}}
\newcommand{\vv}{\mathbf{v}}
\newcommand{\ww}{\mathbf{w}}
\newcommand{\rr}{\mathbf{r}}
\renewcommand{\div}{\mathrm{div}} 
\newcommand{\Hess}{\mathrm{Hess}}
\newcommand{\eps}{\varepsilon}
\newcommand{\zzeta}{\boldsymbol{\zeta}}
\newcommand{\dive}{\mathrm{div}}
\newcommand{\DD}{\mathbf{D}}
\newcommand{\BB}{\mathbf{B}}
\newcommand{\EE}{\mathbf{E}}
\newcommand{\HH}{\mathbf{H}}
\newcommand{\dd}{\mathrm{d}}
\newcommand{\JJ}{\mathbf{J}}
\newcommand{\DMS}{\mathfrak{D}}
\newcommand{\mue}{{\tilde{\mu}}}
\newcommand{\mol}[1]{\bar{#1}}
\newcommand{\const}{\mathrm{const}}

\newcommand{\suma}{\sum_\alpha}

\newcommand{\sumb}{\sum_\beta}

\newcommand{\intd}{\int\,\mathrm{d}}
\newcommand{\ra}{{\rho_\alpha}}

\newcommand{\rac}{{\rho^\dagger_\alpha}}
\newcommand{\racs}{{\rho^*_\alpha}}
\newcommand{\sa}{{s_\alpha}}
\newcommand{\sac}{{s^\dagger_\alpha}}
\newcommand{\sacs}{{s^*_\alpha}}
\newcommand{\ua}{{\uu_{\alpha}}}
\newcommand{\uac}{{\uu^\dagger_{\alpha}}}
\newcommand{\uacs}{{\uu^*_{\alpha}}}

\newcommand{\ub}{{\uu_{\beta}}}
\newcommand{\ubc}{{\uu^\dagger_{\beta}}}
\newcommand{\uai}{{u_{\alpha\,i}}}
\newcommand{\uaci}{{u^{\dagger\, i}_{\alpha}}}
\newcommand{\uaj}{{u_{\alpha\,j}}}
\newcommand{\uacj}{{u^{\dagger\, j}_{\alpha}}}

\newcommand{\uE}{{^\uparrow}E}
\newcommand{\dE}{{^\downarrow}E}
\newcommand{\num}{\textrm{num}}

\begin{document}

\title{A multiscale thermodynamic generalization of Maxwell-Stefan diffusion equations and of the dusty gas model}
\author[wias]{Petr Vágner\corref{cor1}}
\ead{petr.vagner@wias-berlin.de}
\author[mff]{Michal Pavelka}
\author[wias]{Jürgen Fuhrmann}
\author[fjfi]{Václav Klika}


\cortext[cor1]{Corresponding author}

\address[wias]{Weierstrass Institute, Mohrenstr. 39, 10117 Berlin, Germany}
\address[mff]{Mathematical Institute, Faculty of Mathematics and Physics, Charles University in Prague, Sokolovská 83, 186 75 Prague, Czech Republic}
\address[fjfi]{Department of Mathematics, FNSPE, Czech Technical University in Prague, Trojanova 13, Prague 2, 120 00, Czech Republic}

\begin{abstract}
Despite the fact that the theory of mixtures has been part of non-equilibrium thermodynamics and engineering for a long time, it is far from complete.
While it is well formulated and tested in the case of mechanical equilibrium (where only diffusion-like processes take place), the question how to properly describe homogeneous mixtures that flow with multiple independent velocities that still possess some inertia (before mechanical equilibrium is reached) is still open.
Moreover, the mixtures can have several temperatures before they relax to a common value. In this paper, we derive a theory of mixtures from Hamiltonian mechanics in interaction with electromagnetic fields.
The resulting evolution equations are then reduced to the case with only one momentum (classical irreversible thermodynamics), 
providing a generalization of the Maxwell-Stefan diffusion equations. 
In a next step, we reduce that description to the mechanical equilibrium (no momentum) and derive a non-isothermal variant of the dusty gas model. 
These reduced equations are solved numerically, and we illustrate the results on efficiency analysis, showing where in a concentration cell efficiency is lost. Finally, the theory of mixtures identifies the temperature difference between constituents as a possible new source of the Soret coefficient. For the sake of clarity, we restrict the presentation to the case of binary mixtures; the generalization is straightforward.
\end{abstract}


\maketitle

\numberwithin{equation}{section}

\section{Introduction}

Transport models are a hallmark application of non-equilibrium thermodynamics (NET), backed up by a vast experimental evidence for the applicability of linear closures between thermodynamic fluxes and forces \cite{lebon-understanding}. But should one consider a single or multiple constituents for a plausible description of the transport? And how many independent velocities should be present in the models? And what if temperatures of the constituents are not the same \cite{chen}?

Let us pause on the famous Duncan-Toor experiment where a rather unexpected behaviour of diffusion was observed if the mixture is seen as a single continuum. 
Various ``regimes'' of diffusion are reported which are not found in the classical Fickian diffusion, where the diffusive flux occurs solely in the direction of the negative chemical potential gradient, see \cite{krishna-wesselingh, Klika2021}. However, if one just considers the same problem as a mixture, linear nonequilibrium thermodynamics offers a resolution of these new phenomena by very standard means, namely by coupling among the well-known driving forces of transport. This model is known as Maxwell-Stefan model of diffusion \cite{kjelstrup}. Hence, there are potentially key phenomena in transport that can be captured only on the detailed mixture level. On one hand, Maxwell-Stefan models, as well as other transport models, can be viewed as a special case of a more general NET framework, the mixture theory \footnote{We refer the interested reader to reviews \cite{AtkinCraine76,bedford1983theories,venerus2018modern, vasek-review} and to classical texts \cite{bowen1976theory, deBoerbook, dgm, drewPassman2006theory, gray2014introduction, gurtin2010mechanics,rajagopal1995mechanics, Truesdell84}.}. On the other hand, there is a recognised distinguished limit of the Maxwell-Stefan model, the dusty gas model of porous media transport.


We can order the various transport theories as follows:
\begin{itemize}
\item no coupling of thermodynamic forces (no friction, i.e. transfer of momentum, among the constituents), isothermal case -- Fick's law of transport
\item coupling among thermodynamic driving forces for friction, isothermal case -- Maxwell-Stefan model
  \item coupling among thermodynamic driving forces, nonisothermal case -- this article.
  \end{itemize}
Moreover, in this paper we go beyond by adding further state variables to the classical densities and temperature, namely 
\begin{itemize}
\item electromagnetic fields, 
\item one or two velocities \footnote{we restrict ourselves to two component mixtures without loss of generality}
\item and possibly also two temperatures.
\end{itemize}
In order to derive such complex theories, however, we start with the most detailed level, which can be derived from the Liouville equation \cite{PKG}, and then we reduce it by letting the additional state variables relax. 

In particular, we derive a nonisothermal generalisation of the Maxwell-Stefan model that has the overall velocity as an independent state variable, so that the evolution is not purely dissipative. Because we start from the Liouville equation, we obtain also highly non-trivial terms in the evolution equation for momentum density, which can not be found in previous works \cite{bothe-dreyer,pavelka-ijes,soucek2013}. In other words, by starting from the Liouville equation, we reveal mechanical terms that are not visible when one starts from the standard balance equations only. Moreover, the framework leads to a nonisothermal dusty gas model, which we include as an application of the presented theory, as well as to a new origin of the Soret coefficient \cite{kempers2001}.

The paper is organised as follows. Section \ref{sec.gen} contains Hamiltonian and thermodynamic evolution of binary homogeneous mixtures, derived from the microscopic dynamics (Liouville equation), and Hamiltonian vacuum electrodynamics. We rely on the General Equation for Non-Equilibrium Reversible-Irreversible Coupling (GENERIC) \cite{go,og, HCO,PKG}, which combines Hamiltonian mechanics with gradient dynamics. 
Section \ref{sec.em} couples the motion of mixtures with electromagnetic fields in a framework of Galilean invariant electrodynamics of mixtures.
Section \ref{sec.cit} then contains a reduction of the description to a model with a single velocity, which is thus in the realm of Classical Irreversible Thermodynamics \cite{dgm}. 
Section \ref{sec.mecheq} further reduces the model to the mechanical equilibrium, where no inertia is present anymore.
Section \ref{sec.dgm} contains a particular application of the reduced model -- a generalization of the dusty gas model.
Section \ref{sec.full} shows an alternative and more precise route towards the reduced models, where a detailed knowledge of the entropy (either in terms of partial entropies or in terms of partial temperatures) is not necessary.
Finally, Section \ref{sec.soret} shows the implication for the modeling of the Soret coefficient.

\section{GENERIC evolution of mixtures}\label{sec.gen}
Our goal in this section is to formulate models for a mixture at various levels, with various levels of details, but in a systematic and thermodynamically consistent way. To this end, we shall formulate our approach within GENERIC framework \cite{go,og, HCO, PKG}, which combines Hamiltonian mechanics (generated by a Poisson bracket and energy) and gradient dynamics (generated by a dissipation potential and entropy).
Apart from its convenience for multiscale problems, GENERIC clearly identifies and separates reversible and irreversible processes in the evolution of state variables. We shall start by the former.

\subsection{Hamiltonian mechanics of mixtures}
The reversible Hamiltonian evolution of the mixture is generated by a Poisson bracket and energy (or free energy due to the assumption of isothermality), as shown for example in \cite{hierarchy}. 
The Poisson bracket is derived from the Liouville equation \cite{PKG} and reads
\begin{equation}\label{eq.PB}
    \{A,B\} = \suma \{A,B\}^{(FM)_\alpha}
    ~,
\end{equation}
where $A$ and $B$ are two arbitrary sufficiently regular functionals of the state variables: partial density $\rho_\alpha$, partial momentum density $\uai$, and partial entropy density $\sa$. The fluid mechanics Poisson bracket for species $\alpha$, that is the bracket corresponding to the fluid mechanics level for each individual species, is given by 
\begin{eqnarray}
 \{A,B\}^{(FM)_\alpha} &=& \intd\rr \rho_\alpha \left(\partial_i A_\ra B_\uai - \partial_i B_\ra A_\uai\right)\nonumber\\
    &&+ \intd\rr \uai \left(\partial_j A_{\uai} B_{\uaj} - \partial_i B_{\uai} A_{\uaj}\right)\nonumber\\
 &&+ \intd\rr \sa \left(\partial_j A_\sa B_\uaj - \partial_i B_\sa A_\uaj\right)
 ~.
\end{eqnarray}
Note that for instance expression $A_\ra$ stands for functional derivative of functional $A$ with respect to the field $\ra(t,\rr)$, $\frac{\delta A}{\delta \ra(t,\rr)}$, see for instance \cite{HCO,PKG}, and $\partial_i$ stands for spatial derivative $\frac{\partial}{\partial r^i}$. Einstein summation of spatial indexes is employed in the whole manuscript.


Given total energy of the system $E$, reversible evolution of an arbitrary functional $A$ can be then expressed in two ways, see \cite{PKG},
\begin{subequations}
\begin{eqnarray}
\label{eq.dA.PB} \left(\frac{\mathrm{d} A}{\mathrm{d} t}\right)_{rev} &=& \{A,E\}\\
\label{eq.dA.dx} &=& \intd\rr \suma A_\ra \frac{\partial \ra}{\partial t} + \suma A_\ua \cdot \frac{\partial \ua}{\partial t}+ \suma A_\sa \cdot \frac{\partial \sa}{\partial t}.
\end{eqnarray}
\end{subequations}
By rewriting expression \eqref{eq.dA.PB} in the form of Eq. \eqref{eq.dA.dx}, we can read the evolution equations of $\ra$, $\ua$ and $\sa$,
\begin{subequations}\label{eq.evo.rev.c}
\begin{eqnarray}
\left(\frac{\partial \ra}{\partial t}\right)_{rev} &=& -\partial_i(\ra \uaci)~,\\
\left(\frac{\partial \uai}{\partial t}\right)_{rev} &=& -\ra\partial_i\rac -\uaj \partial_i \uacj -\partial_j\left(\uai \uacj\right)~,\\
\left(\frac{\partial \sa}{\partial t}\right)_{rev} &=& -\partial_i(\sa \uaci)\\
\end{eqnarray}
where the conjugate densities, $\rac$, and conjugate momenta, $\uac$, and conjugate entropies, $\sac$, are identified with the respective derivatives of energy, 
\begin{equation}
 \rac = E_\ra,  \uac = E_\ua \mbox{ and } \sac = E_\sa~.
\end{equation}
\end{subequations}

\subsection{Hamiltonian evolution of electromagnetic field in vacuum}
The reversible evolution of the displacement field $\DD$ and magnetic field $\BB$ in vacuum is generated by the Poisson bracket
\begin{eqnarray}
\{A,B\}^{(\textrm{EM})} = \int A_{D^i} \varepsilon^{ijk} \partial_j B_{B^k} - B_{D^i} \varepsilon^{ijk}\partial_j A_{B^k} 
~,
\label{eq.EMPB}
\end{eqnarray}
where we denoted with $\varepsilon^{ijk}$ the Levi-Civita symbol. 
For an arbitrary energy E, the evolution equation of $\DD$ and $\BB$ read
\begin{subequations}
\begin{eqnarray}
\partial_t D^i &=   \varepsilon^{ijk} \partial_j E_{B_k}  
~,
\\
\partial_t B^i&= -  \varepsilon^{ijk} \partial_j E_{D_i}
~,
\end{eqnarray}
where $E_{\DD} = \EE$ is the electric field and $E_{\BB}=\HH$ is the magnetic intensity, see~\cite{AffroCMAT}, \cite{PKG}. 
\end{subequations}

\subsection{Irreversible evolution}
Within the used GENERIC framework, irreversible evolution of the mixture is generated by a dissipation potential \cite{pre15}. 
Namely
\begin{subequations}
 \begin{eqnarray}
\left(\frac{\partial \ra}{\partial t}\right)_{irr} &=& \frac{\delta \Xi}{\delta \racs}~,\\
\left(\frac{\partial \ua}{\partial t}\right)_{irr} &=& \frac{\delta \Xi}{\delta \uacs}~,\\
\left(\frac{\partial \sa}{\partial t}\right)_{irr} &=& \frac{\delta \Xi}{\delta \sacs}~,
 \end{eqnarray}
where the conjugates with respect to entropy appear
    \begin{equation}
 \racs = S_\ra,  \uacs = S_\ua \mbox{ and } \sacs = S_\sa.
\end{equation}
We assume the simplest possible form of dissipation potential \cite[Ch 6]{PKG} being algebraic and quadratic in the conjugate variables
\begin{equation}\label{eq.Xi}
\Xi 
=
\frac{1}{2}
\intd\rr 
	\frac{1}{\frac{1}{n}\sum_\gamma s_\gamma^\dagger}
	\frac{1}{2} 	
	\sum_{\stackrel{\alpha,\beta=1}{\beta\neq\alpha}}^n  
	\zeta_{\alpha\beta}  (\ua^\dagger-\ub^\dagger)^2	
	+
	\frac{1}{\frac{1}{n}\sum_\gamma s_\gamma^\dagger} 
	\frac{1}{2} 
	\sum_{\stackrel{\alpha,\beta=1}{\beta\neq\alpha}}^n  
	K_{\alpha\beta}  (s_\alpha^\dagger-s_\beta^\dagger)^2
\end{equation}
where the prefactor $\frac{1}{n}\sum_{\alpha=1}^n s_\alpha^\dagger$ corresponds to the average temperature of the mixture ($s^\dagger$, the conjugate of the total entropy $s=\suma s_a$), as will be shown below.

Using Eq. \eqref{eq.Xi}, the irreversible evolution becomes explicitly
\end{subequations}
\begin{subequations}\label{eq.evo.irr}
 \begin{align}
\left(\frac{\partial \ra}{\partial t}\right)_{irr} &= 0\\
\left(\frac{\partial \ua}{\partial t}\right)_{irr} &= \sumb \zeta_{\alpha\beta}(\ubc-\uac) \\
\left(\frac{\partial \sa}{\partial t}\right)_{irr} &= \frac{1}{\sum_{\gamma=1}^n s_\gamma^\dagger} \left[\sum_{\beta,\gamma} \zeta_{\beta\gamma}(\ubc-\uu_\gamma^\dagger)^2
+ \sum_{\beta\neq\alpha} K_{\alpha\beta} (s_\beta^\dagger - s_\alpha^\dagger)(s_\beta^\dagger + \sum_{\gamma\neq\alpha} s_\gamma^\dagger)
\right.\nonumber\\
&
\phantom{sssssssssssss}
\left.+  \frac{1}{2}\sum_{\stackrel{\beta,\gamma=1}{\beta,\gamma\neq\alpha}}^n K_{\beta\gamma}(s_\beta^\dagger-s_\gamma^\dagger)^2\right]
 \end{align}
\end{subequations}
The zero-th component was separated as it will be considered as the dust in Sec. \ref{sec.dgm}.

Coefficients $\zeta_{\alpha\beta}$ are to be interpreted as mutual friction coefficients as in \cite{Bearman1961}, and they are assumed to be symmetric, i.e. $\zeta_{\alpha\beta} = \zeta_{\beta\alpha}$. This symmetry can be regarded as the validity of Onsager reciprocal relations \cite{onsager1930,onsager1931,PRE2014} which has been recognised and accepted in particular in transport problems \cite{klika2018beyond, klika2017functional}.
Similarly, the heat transfer coefficients $K_{\alpha\beta}$ are also assumed to be symmetric, i.e.~$K_{\alpha\beta} = K_{\beta\alpha}$. 
In the following section, we add coupling between mechanics of mixtures and electrodynamics. 

\section{Binary electromagnetohydrodynamics with friction}\label{sec.em}
Since charged matter within electromagnetic fields is subject to the Lorentz force, we have to combine the building blocks introduced in the previous section to formulate the evolution equations of the binary electromagnetohydrodynamic mixture with friction. 
First, we couple the fluid mechanics Poisson bracket~\eqref{eq.PB} with the electromagnetic Poisson bracket~\eqref{eq.EMPB} using the semi-direct product technique, see~\cite{AffroCMAT}.
Second, we endow this coupled Poisson bracket with the irreversible terms generated by the dissipation potential~\eqref{eq.Xi}. 
We present the resulting evolution equations below. 



\subsection{Evolution equations}
Evolution equations describing a binary charged mixture read
\begin{subequations}\label{eq.EMHD2}
\begin{align}
  \partial_t \rho_a   
  &= 
  -\partial_i\left(\rho_a v_a^i \right)
  ~,
  \\
	\label{eq.Euler2.ua} 
  \partial_t u_{a,i}    
  &= 
  -\rho_a \partial_i \frac{\partial \eps}{\partial \rho_a} 
  - s_a \partial_i \frac{\partial \eps}{\partial s_a}
  -\partial_j\left(u_{a,i} v_a^j \right)
  \nonumber
  \\
  &+\rho_a \frac{q_a}{m_a} 
  \left(
    E_i + \epsilon_{ijk}v_a^j B^k
  \right) 
  +\zeta_{ab}(v_b^i-v_a^i)
  ~,
  \\
  \partial_t s_a      
  &= 
  -\partial_i\left( s_a v_a^i \right) 
   +\frac{1}{T_a + T_b}
   	\left(
		\zeta_{ab}(\uu_b^\dagger-\uu_a^\dagger)^2
		+
		2K_{ab}T_b(T_b - T_a)
	\right)
  ~,
  \\
  \partial_t \rho_b   
  &= 
  -\partial_i\left(\rho_b v_b^i \right)
  ~,
  \\
	\label{eq.Euler2.ub} 
  \partial_t u_{b,i}    
  &= 
  -\rho_b \partial_i \frac{\partial \eps}{\partial \rho_b} 
  -s_b \partial_i \frac{\partial \eps}{\partial s_b}
  -\partial_j\left(u_{b,i} v_b^j \right)
  \nonumber
  ~,
  \\
  &+\rho_b \frac{q_b}{m_b} \left(E_i + \epsilon_{ijk}v_b^j B^k\right) 
  + \zeta_{ab}(v_a^i-v_b^i)
  ~,\\
  \partial_t s_b      
  &= 
  -\partial_i\left( s_b v_b^i \right) 
+ \frac{1}{T_a+T_b} \left(
		\zeta_{ab}(\uu_b^\dagger-\uu_a^\dagger)^2 
 		+
		2K_{ab}T_a(T_a - T_b)
	\right)
  ~,
  \\
  \partial_t D^i 
  &= 
  \epsilon^{ijk}\partial_j H_k 
  -\left(\frac{q_a \rho_a}{m_a} v_a^i 
  +\frac{q_b \rho_b}{m_b} v_b^i\right)
  ~,
  \\
  \partial_t B^i 
  &= 
  -\epsilon^{ijk}\partial_j E_k
  ~,
\end{align}
\end{subequations}
where we denoted electric $\EE$ and magnetic intensities $\HH$, and electric $\DD$ and magnetic field $\BB$, internal energy $\eps$ and charge flux $q_\alpha$  \cite{PKG} for more details and relation to the kinetic theory. By integration of the equations for $\DD$ and $\BB$ with respect to time, the static Maxwell equations are obtained,
\begin{align}\label{eq.EMHD2.divDB}
\dive \DD = \frac{q_a\rho_a}{m_a}+\frac{q_b\rho_b}{m_b}
\qquad\mbox{and}\qquad
\dive \BB = 0.
\end{align}
Note that the total energy is conserved and the total entropy density,
\begin{align}
s = s_a + s_b,
\end{align}
grows. 
Volumetric total energy density can be prescribed, for example, as 
\begin{align}
    \label{eq.e.EMHD2}
    e &= \frac{1}{2}\rho_a\vv_a^2 + \frac{1}{2}\rho_b\vv_b^2 
                  + \eps(\rho_a, \rho_b ,s_a, s_b) 
		  +\frac{1}{2\epsilon}\DD^2 + \frac{1}{2\mu}\BB^2 + \frac{\rho_a \vv_a +\rho_b \vv_b}{\rho_a+\rho_b} \cdot(\DD\times\BB)
~,
\end{align}
so that it consists of the total kinetic energy density, internal energy and we have $\EE = \frac{\partial e}{\partial \DD}$, $\HH = \frac{\partial e}{\partial \BB}$, see \cite{AffroCMAT}. In addition, we adopt the standard notation for $\mu_{a,b} =\frac{\partial \eps}{\partial \rho_{a,b}}$, $T_{a,b} =\frac{\partial \eps}{\partial s_{a,b}}$, and $\mm = \frac{\partial e}{\partial \vv_a}+\frac{\partial e}{\partial \vv_b}$ for the chemical potentials, partial temperatures, and mass momentum, respectively.

The mixture is equipped with two entropies $s_a$ and $s_b$. That means that the mixture has also two temperatures $T_a$ and $T_b$. This situation is usual in low-temperature plasma, where electrons have higher temperature than ions because they absorb energy of rapidly varying electromagnetic field more efficiently~\cite{chen}. 

\subsection{Two-temperature local equilibrium thermodynamics}
When summing the two equations for $\uu_a$ and $\uu_b$, one can read the total pressure (noting that pressure follows from the definition of the state variables and energy and hence is not subjected to any constitutive choices),
\begin{align}
-p = -\epsilon + \rho_a \mu_a + \rho_b \mu_b + s_a T_a + s_b T_b, 
\end{align}
and the Gibbs-Duhem relation
\begin{align}
-\dd p =  \rho_a \dd \mu_a + \rho_b \dd \mu_b + s_a \dd T_a+ s_b\dd T_b
\end{align}
or
\begin{align}\label{eq.EMHD2.GD}
-\nabla p =  \rho_a \nabla \mu_a + \rho_b \nabla \mu_b + s_a \nabla T_a+ s_b\nabla T_b.
\end{align}

Differential of the local equilibrium energy (standing for the internal energy per a small volume) can then be written as
\begin{align}
\dd E = T_a \dd S_a + T_b \dd S_b - p \dd V + \mu_a \dd M_a + \mu_b \dd M_b,
\end{align}
the differential of the Gibbs free energy $G= E-T_a S_a - T_b S_b + pV$ becomes
\begin{align}
\dd G = - S_a \dd T_a  -S_b \dd T_b + V \dd p + \mu_a \dd M_a + \mu_b \dd M_b.
\end{align}
Therefore, we obtain the Maxwell relations
\begin{align}\label{eq.EMHD2.sa.mu}
s_a = \rho_a \left(\frac{\partial S_a}{\partial M_a}\right)_{T_a, T_b, p, M_b} =
-\rho_a\left(\frac{\partial \mu_a}{\partial T_a}\right)_{T_b, p, M_a, M_b}.
\end{align}

This definition of $s_a$ is indeed compatible with the foregoing definition, i.e. the local volumetric entropy density.
The Euler theorem for 1-homogeneous functions implies for the total entropy, which is extensive, that 
\begin{align}
S = S_a + S_b 
\qquad\mbox{for}\qquad
S_a =s_a \frac{M_a}{\rho_a},  S_b = s_b \frac{M_b}{\rho_b}.
\end{align}
Therefore, $s_a$ defined in \eqref{eq.EMHD2.sa.mu} represent volumetric densities of entropies of the species (when assuming no entropy attributed to the microscopic interfaces between constituents).

Moreover, we can define the partial enthalpy  by 
\begin{align}
	h_\alpha= \mu_\alpha + T\frac{s_\alpha}{\rho_\alpha} = \mu_\alpha + T\left(\frac{\partial S_\alpha}{\partial M_\alpha}\right)_{T,p,M_{\beta\neq\alpha}},
\end{align}
which plays a role in the energy conservation below. In the next Section, we reduce the rather complicated models from the current Section to less detailed descriptions.

\section{Binary electrohydrodynamics within Classical Irreversible Thermodynamics}\label{sec.cit}
Let us, in two steps, reduce model~\eqref{eq.EMHD2}  to a less detailed model on the level of the Classical Irreversible Thermodynamics (CIT)~\cite{dgm}, that is to a description with only one velocity. 
In the first step, the temperatures of the constituents $T_a$ and $T_b$ will be equalized to a single temperature $T_a=T=T_b$ using the MaxEnt principle. 
In the second step, the partial momenta $\ua$ and $\ub$ will be approximated to the total momentum $\uu = \ua + \ub$ and the diffusion velocity $\ww^\dagger$ using the Dynamic MaxEnt.
However, before proceeding to the reductions, we let for simplicity the magnetic field relax to $\BB=0$.

\subsection{MaxEnt reduction to one-temperature continuum}\label{sec.reduced-entropies}
In the two-temperature treatment we were equipped with two entropies, $s_a$ and $s_b$. After the two temperatures have locally relaxed to their common value $T$, we are equipped with only one entropy $s=s_a+s_b$. What is the least-biased estimate of the partial entropies $s_a$ and $s_b$ in that case? We shall use the MaxEnt principle~\cite{RedExt,PKG,dynmaxent}.

Since we are working with entropies as state variables, it is more comfortable to minimize energy rather than maximize entropy.\footnote{Both approaches are equivalent as far as we have entropy or energy as a state variable \cite{callen}. They are not equivalent for instance in kinetic theory, where neither energy nor entropy are among the state variables and where only the distribution function plays the role of state variable \cite{PKG}.} 
Consider a detailed energy $\uE(s_a, s_b, \rho_a, \rho_b, \uu_a, \uu_b)$. 
The reduction to the total entropy $s=s_a+s_b$ is carried out by the reducing Legendre transformation, see \cite{RedExt},
\begin{align}
    \frac{\partial}{\partial s_a}\left(-\uE + \langle T, s_a +s_b\rangle\right) = 0 
    =\frac{\partial}{\partial s_b}\left(-\uE + \langle T, s_a +s_b\rangle\right),
\end{align}
which is equivalent to the equations
\begin{align}\label{eq.sab.T}
    \frac{\partial \uE}{\partial s_a} = T = \frac{\partial \uE}{\partial s_b},
\end{align}
solution of which are the dependencies $\tilde{s}_a(T, \rho_a, \rho_b)$ and $\tilde{s}_b(T, \rho_a, \rho_b)$, which we will further refer to as \emph{reduced partial entropy densities}. Note that these functions can be explicitly obtained only after an explicit formula for $\uE$ has been provided. 

We can proceed in the Legendre transformation by defining the negative Helmholtz free energy
\begin{align}
    \dE^*(T, \rho_a, \rho_b,\uu_a,\uu_b) = -\uE(\tilde{s}_a(T, \rho_a, \rho_b), \tilde{s}_b(T, \rho_a,\rho_b), \rho_a, \rho_b,\uu_a,\uu_b) + T\cdot(\tilde{s}_a(T, \rho_a, \rho_b) + \tilde{s}_a(T, \rho_a, \rho_b)).
\end{align}
Finally, the Legendre transformation (this time not reducing, but invertible)
\begin{align}
    \frac{\partial}{\partial T}\left(-\dE^* + T s\right) = 0,
\end{align}
which leads to a dependency $T(s, \rho_a, \rho_b,\uu_a,\uu_b)$ and the final formula for the reduced energy
\begin{align}
    \dE(s, \rho_a, \rho_b,\uu_a,\uu_b) = -\dE^*(T(s, \rho_a, \rho_b), \rho_a, \rho_b,\uu_a,\uu_b) + T(s, \rho_a, \rho_b,\uu_a,\uu_b) s.
\end{align}

Differential of the free energy reads
\begin{align}
    d\dE^* = -s dT + \mu_a d\rho_a + \mu_b d\rho_b+v_a^i d u_a^i+v_b^i d u_b^i, 
\end{align}
which leads to the Maxwell relations
\begin{align}
    \left(\frac{\partial s}{\partial \rho_{a,b}}\right)_{T,\uu_a,\uu_b} = -\left(\frac{\partial \mu_{a,b}}{\partial T}\right)_{\rho_a,\rho_b,\uu_a,\uu_b}.
\end{align}
Recalling that $s=s_a + s_b$, we can write, for instance, that
\begin{align} \label{eq.sa.gen}
    \left(\frac{\partial s_a }{\partial \rho_{a}}\right)_T + \left(\frac{\partial s_b }{\partial \rho_{a}}\right)_T = -\left(\frac{\partial \mu_a}{\partial T}\right)_{\rho_a,\rho_b},
\end{align}
for fixed partial momenta $\uu_a,~\uu_b$. If the second term on the left hand side vanishes, as in the case of non-interacting ideal gases \cite{PKG}, we obtain 
\begin{align}\label{eq.sa.T.mu}
    \left(\frac{\partial s_a }{\partial \rho_{a}}\right)_T = -\left(\frac{\partial \mu_a}{\partial T}\right)_{\rho_a,\rho_b},
\end{align}
which will become useful later on. Note however, that this formula ceases to hold in case of interacting gases and one has to rely on \eqref{eq.sa.gen}.

\subsection{DynMaxEnt reduction of the relative momentum $\ww$}
In order to reduce the dynamical equations~\eqref{eq.EMHD2}, we first rewrite the two partial momenta into the total momentum $\uu$ and relative momentum $\ww$, as follows,  
\begin{subequations}
\begin{align}
\uu &= \uu_a + \uu_b
~,\\
\ww &= \frac{1}{2}(\uu_a - \uu_b)
~,
\end{align}
and conversely,
\begin{align}
\uu_a &= \frac{\uu}{2}+\ww
~,\\
\uu_b &= \frac{\uu}{2}-\ww.
\end{align}
\end{subequations}
The corresponding conjugate variables with respect to energy 
\begin{subequations}
\begin{align}
\uu^\dagger &= \frac{1}{2}(\uu^\dagger_a + \uu^\dagger_b)
~,\\
\ww^\dagger &= \uu^\dagger_a - \uu^\dagger_b
~,
\end{align}
and conversely,
\begin{align}
\uu^\dagger_a &= \uu^\dagger + \frac{1}{2}\ww^\dagger
~,\\
\uu^\dagger_b &= \uu^\dagger - \frac{1}{2}\ww^\dagger
~.
\end{align}
Note that the conjugate to the total momentum can be interpreted as the barycentric velocity, $\uu^\dagger = \vv$.
\end{subequations}

In order to identify the reduced version of dynamics, we have to specify the state variables on the lower (less-detailed) level together with its link to the upper (more-detailed) level state variables. Consequently, entropy and MaxEnt principle ~\cite{dynmaxent} provides the least biased estimate of the reduced dynamics.

Our aim is to identify the dynamics of the system with a single macroscopic velocity/momentum and temperature. Hence, the lower state variables are $(\rho_a,\rho_b, \uu,s)$. The MaxEnt estimate of the relative momentum $\ww$ (while keeping the total momentum unchanged) reads
\begin{subequations}
    \label{eq.w.MaxEnt}
\begin{align}
 &\uu_a = \frac{\rho_a}{\rho_a+\rho_b} \uu~, \qquad &\uu_b  = \frac{\rho_b}{\rho_a+\rho_b}\uu 
 ~,\\
 &\uu = \uu~, \qquad & \ww = \frac{\rho_a-\rho_b}{2(\rho_a+\rho_b)} \uu
 ~,
\end{align}
\end{subequations}
see e.g. \cite{PKG}, which we shall substitute into the equations. The conjugate variable $\ww^\dagger$ is to be determined as the solution to the stationary evolution equation for $\ww$~\cite{dynmaxent}. Finally after the reduction, Eqs.~\eqref{eq.EMHD2} become
\begin{subequations}\label{eq.EMHD2.CIT}
\begin{align}
\label{eq.EMHD2.CIT.rho_a}
\partial_t \rho_a   
&= 
-\partial_i\left(\rho_a \left(v^i +\frac{w^{\dagger i}}{2}\right)\right)
~,
\\
\label{eq.EMHD2.CIT.rho_b}
\partial_t \rho_b   
&= 
-\partial_i\left(\rho_b \left(v_i -\frac{w^{\dagger i}}{2}\right)\right)
~,
\\
\partial_t u_i  \label{eq.EMHD2.CIT.u}
&= 
-\rho_a \partial_i \mu_a 
- \tilde{s}_a\partial_i T 
-\rho_b \partial_i \mu_b 
- \tilde{s}_b\partial_i T 
\\
&-\partial_j\left(
  \frac{\rho_a}{\rho_a+\rho_b} u_i \left(v^j+\frac{w^{\dagger j}}{2}\right) 
  +\frac{\rho_b}{\rho_a+\rho_b} u_i \left(v^j-\frac{w^{\dagger j}}{2}\right) 
\right)
+\left(\rho_a \frac{q_a}{m_a}+\rho_b \frac{q_b}{m_b}\right) E_i  
\nonumber
\\
&= -\rho_a \partial_i \mu_a 
- \tilde{s}_a\partial_i T 
-\rho_b \partial_i \mu_b 
- \tilde{s}_b\partial_i T
-\partial_j
\left( 
  u_i v^j 
  +\frac{\rho_a-\rho_b}{2(\rho_a+\rho_b)}u_i w^{\dagger j}
\right)
+\left(\rho_a \frac{q_a}{m_a}+\rho_a \frac{q_b}{m_b}\right) E_i 
\nonumber
~,
\\
\partial_t s  
&= 
-\partial_i
\left( 
  \tilde{s}_a \left(v^i+w^{\dagger i}/2\right) 
  +\tilde{s}_b \left(v^i-w^{\dagger i}/2\right)
\right) 
+\frac{1}{2 T} \zeta_{ab} (\mathbf{v}_a-\mathbf{v}_b)^2~,
\\
%
%
\label{eq.EMHD2.CIT.w}
\zeta_{ab} w^{\dagger i}
 &= 
    \frac{1}{2}\left(
       -\rho_a\partial_i\mu_a - \tilde s_a\partial_i T + \frac{\rho_a}{m_a}q_a E_i
       +\rho_b\partial_i\mu_b + \tilde s_b\partial_i T - \frac{\rho_b}{m_b}q_b E_i
    \right)
   +\frac{1}{4}\partial_j\left(
        u_i w^{\dagger j}
    \right)
\\
\partial_t D^i &=-\left(\frac{q_a \rho_a}{m_a} v_a^i +\frac{q_b \rho_b}{m_b} v_b^i\right),
    \end{align}
where the total entropy $s=\tilde{s}_a+\tilde{s}_b$ and the reduced partial entropies $\tilde{s}_{a,b}(T)$ are defined as functions of temperature $T$; see Eqs. \eqref{eq.sab.T}. We refer the reader to the section~\ref{sec.sa_example} where we work out an example of the reduced entropies $\tilde{s}_{a,b}$ . 
\end{subequations}
Neither the term in the equation for $\uu$ that depends on $\ww^\dagger$, nor the terms in the equation for $\ww^\dagger$ that depend on $\uu$ could be seen without starting from the Liouville equation, which shows the advantages of our approach to the standard approach based on balance equations \cite{bothe-dreyer}.

\subsubsection{Approximate diffusion velocity $\widehat{\ww}^\dagger$}
The velocity ${\ww}^\dagger$ appears with alternating signs in the continuity equations~\eqref{eq.EMHD2.CIT.rho_a} and~\eqref{eq.EMHD2.CIT.rho_b} besides the barycentric velocity $\vv=\uu^\dagger$. It is thus tempting to interpret $\ww^\dagger$ as a diffusion velocity. How can one see that? To this end, let us neglect the divergence term $\partial_j (u_i w^{\dagger j})$ in equation~\eqref{eq.EMHD2.CIT.w}. We use the remaining terms to define the approximate velocity,
\begin{align}
    \widehat{w}^{\dagger i} :=  
    \frac{1}{2\zeta_{ab}}\left(
       -\rho_a\partial_i\mu_a - \tilde s_a\partial_i T + \frac{\rho_a}{m_a}q_a E_i
       +\rho_b\partial_i\mu_b + \tilde s_b\partial_i T - \frac{\rho_b}{m_b}q_b E_i
    \right)
    \label{eq.approx-wd}
    ~.
\end{align}
Clearly, the velocity $\widehat{\ww}^\dagger$ is proportional to the difference of the thermodynamic forces\footnote{gradients of the chemical potential and temperature, and electric field} acting upon the mixture species. Hence, together with its role in the continuity equations,~\eqref{eq.EMHD2.CIT.rho_a} and~\eqref{eq.EMHD2.CIT.rho_b}, $\widehat{\ww}^\dagger$ is the diffusion velocity. This results is consistent with the canonical CIT formulation~\cite{dgm}.

This observation that $\ww^\dagger$ corresponds to diffusion velocity is in line with the definition of the direct variable $\ww$ itself. In particular, $\uu_\alpha^\dagger$ is the partial velocity as can be seen from the standard choice of energy above, \eqref{eq.e.EMHD2}, and the conjugate variables are related via $\uu_\alpha^\dagger = \uu^\dagger\pm\frac{1}{2}\ww^\dagger$ and hence, as $\uu^\dagger$ is the barycentric velocity, we expect the $\ww^\dagger$ variable to represent the diffusion velocity.

However, in contrast to the usual formulations of CIT, such as~\cite{dgm}, the diffusive velocity appears in the evolution equation of the total momentum~\eqref{eq.EMHD2.CIT.u}. What is its role therein? Let us 
integrate the momentum equation~\eqref{eq.EMHD2.CIT.u} over a volume $V$. We get
\begin{align}
\partial_t \int_V u_i 
&= 
\int_{\partial V} -\left( 
    -\varepsilon 
    +\rho_a \mu_a
    +\rho_b \mu_b
    + (\widetilde{s}_a +\widetilde{s}_b)T
\right)\nu_i
+\int_V \left(
  \rho_a \frac{q_a}{m_a}+\rho_a \frac{q_b}{m_b}
\right) E_i
\nonumber
\\
&\phantom{--}
-\int_{\partial V}
u_i\left( 
  v^j 
  +\frac{\rho_a-\rho_b}{2(\rho_a+\rho_b)} \widehat{w}^{\dagger j}
\right) \nu_j
    \label{}
    ~,
\end{align}
where $\nu_i$ denote components of the outer normal. The first integral on the right hand side represents the material pressure; the second expresses the force acting upon $V$ due to the electric field $\EE$; and the last integral is the transport of the total momentum $\uu$ across the boundary $\partial V$. In other words, the total momentum $\uu$ is on top of the (barycentric)-velocity field $\vv$ also advected by the density-weighted diffusion velocity $\frac{\rho_a-\rho_b}{2(\rho_a+\rho_b)} \widehat{\ww}^{\dagger}$. In the case that $\rho_a \gg \rho_b$, for example a dilute aqueous solution of a salt, the $\widehat{\ww}^{\dagger}$ could become in certain flow regimes comparable to $\vv$.

\section{Reduction to mechanical equilibrium}\label{sec.mecheq}
In this Section, we further reduce the description to the level of mechanical equilibrium, where no velocity plays the role of state variable and thus no inertial effects are present.
We start again with system \eqref{eq.EMHD2} and use the MaxEnt single-temperature reduction from Section~\ref{sec.reduced-entropies}.
The partial momenta are put equal to their MaxEnt values, $\uu_a = 0 =\uu_b$; and as it follows from the Dynamic MaxEnt, the partial velocities $\vv_a$ and $\vv_b$ are governed by the reduced momentum balances. As in the previous section, we assume that magnetic field is weak enough, putting $\BB = 0 = \HH$. The above-described reduction of the model~\eqref{eq.EMHD2} reads
\begin{subequations}\label{eq.EMHD2.MechEq}
    \begin{align}
 \label{eq.EMHD2.MechEq.ra}       \partial_t \rho_a   &= -\partial_i\left(\rho_a v_a^i \right)~,\\
 \label{eq.EMHD2.MechEq.rb}       \partial_t \rho_b   &= -\partial_i\left(\rho_b v_b^i \right)~,\\
	    \label{eq.EMHD2.MechEq.ua}0  &= 
        -\rho_a \partial_i \mu_a- \tilde{s}_a\partial_i T + \rho_a \frac{q_a}{m_a} E_i  
+ \zeta_{ab}(v_b^i-v_a^i)\\
	    \label{eq.EMHD2.MechEq.ub}0  &= 
-\rho_b \partial_i \mu_b - \tilde{s}_b\partial_i T+ \rho_b \frac{q_b}{m_b} E_i  
+ \zeta_{ab}(v_a^i-v_b^i)\\
 \label{eq.EMHD2.MechEq.D}\partial_t D^i &= -\left(\frac{q_a \rho_a}{m_a} v_a^i +\frac{q_b \rho_b}{m_b} v_b^i\right)\\
 \label{eq.EMHD2.MechEq.s}       \partial_t s      &= -\partial_i\left( \tilde s_a v_a^i + \tilde s_b v_b^i  \right) + \frac{1}{2 T} \zeta_{ab}(\mathbf{v}_a-\mathbf{v}_b)^2,
    \end{align}
    where $\DD = \epsilon \EE$ and $\EE = -\nabla\varphi$. Again, $\tilde{s}_{a,b}(T)$ are given by Eqs. \eqref{eq.sab.T} and $s=\tilde{s}_a + \tilde{s}_b$. 
\end{subequations}
Note that the two flux-force relations (following from the linear momentum balance) do yield a single equation for the unknown flux $\rho_\alpha \mathbf{v}_\alpha$ while the second relation follows from the condition for mixture mechanical equilibrium. Hence $\rho_\alpha \mathbf{v}_\alpha=-\rho_\beta \mathbf{v}_\beta$ and the sum of \eqref{eq.EMHD2.MechEq}c,d yields Gibbs-Duhem relation. In addition, Galilean invariance has been sacrificed as momentum is no longer among the state variables. The equations are thus valid only in the chosen lab frame, where the total momentum vanishes, that is, $\rho_\alpha \mathbf{v}_\alpha+ \rho_\beta \mathbf{v}_\beta=0$.

In the case of non-interacting gases, we use formula \eqref{eq.sa.T.mu} to rewrite the gradient of chemical potential of species $a$ at constant temperature:
\begin{align}
\nabla(\mu_a)_T &= \left(\nabla\mu_a + \frac{s_a}{\rho_a}\nabla T_a\right)_{T_a=T_b=T} = \nabla\mu_a + \left(\frac{\partial S}{\partial M_a}\right)_{T,p,M_b}\nabla T \nonumber\\
&=\nabla\mu_a - \left(\frac{\partial \mu_a}{\partial T}\right)_{p,M_a,M_b}\nabla T
~,
\end{align}
to further simplify the Eqs. \eqref{eq.EMHD2.MechEq.ua} and \eqref{eq.EMHD2.MechEq.ub},.

Equations \eqref{eq.EMHD2.MechEq.ua} and \eqref{eq.EMHD2.MechEq.ub} can be interpreted as the constitutive relations for the unknown velocities $\vv_a$ and $\vv_b$. Sum of the two equations expresses the momentum balance, using the Gibbs-Duhem relation \eqref{eq.EMHD2.GD},
\begin{equation}
0 = -\nabla p +\left(\rho_a \frac{q_a}{m_a}+\rho_b \frac{q_b}{m_b}\right) \EE.
\end{equation}
The approximation of momentum balance in the mechanical equilibrium is thus implied. 
The total energy density is of the following form $e(\rho_\alpha, \DD,s)=\int \DD^2/(2\epsilon) + \epsilon(T, \rho_a, \rho_b)$.
Entropy is clearly produced in Eqs. \eqref{eq.EMHD2.MechEq} and energy is conserved. Indeed, one obtains by chain rule for the total energy density $e(\rho_\alpha, \DD,s)$ that
\begin{equation}\label{eq.EMHD2.MechEq.e}
	\partial_t e = -\nabla\cdot \JJ_e, \qquad \mbox{for}\qquad \JJ_e = \sum_{\alpha=a,b} \rho_\alpha h_\alpha \vv_\alpha.
\end{equation}
The equations are thus compatible with the standard conservation laws of mass and energy and with the balance of entropy. 

The evolution equation for $\DD$ can be integrated in time to obtain the Poisson equation \eqref{eq.EMHD2.divDB}, or
\begin{equation}\label{eq.EMHD2.Poisson}
-\dive\left(\epsilon \nabla\varphi\right) = \frac{q_a\rho_a}{m_a}+\frac{q_b\rho_b}{m_b},
\end{equation}
which can be solved instead of Eq. \eqref{eq.EMHD2.MechEq.D}. The advantage of having $\DD$ among the state variables is the explicit conservation of energy.
On the other hand, the Poisson equation is more suitable for actual calculation because boundary conditions are typically given by setting the electrostatic potential.

The friction coefficient $\zzeta_{ab}$ should disappear when $\rho_a\rightarrow 0$ or $\rho_b\rightarrow 0$, since then one of the components is missing. Therefore, it is often expressed as 
\begin{align}\label{eq.diffcoef}
\zzeta_{ab} = \frac{k_B T }{\DMS_{ab}}
            \frac{
                \frac{\rho_a}{m_a}
                \frac{\rho_b}{m_b}
            }{
                \frac{\rho_a}{m_a}
                +\frac{\rho_b}{m_b}
            }, 
\end{align}
where $\DMS_{ab}$ is the Maxwell-Stefan interdiffusion coefficient as defined in~\cite{krishna-wesselingh}. 
The advantage is that the interdiffusion coefficients are typically more or less independent of composition \cite{krishna-wesselingh}. 
Equations \eqref{eq.EMHD2.MechEq.ua} and \eqref{eq.EMHD2.MechEq.ub} can be also rewritten as
\begin{align}\label{eq.Maxtef}
\rho_a\frac{\nabla (\mue_\alpha)_T}{k_B T} 
= 
\frac{1}{\frac{\rho_a}{m_a}+\frac{\rho_b}{m_b}}\sum_{\beta\neq\alpha}\frac{\rho_\alpha}{m_\alpha}\frac{\rho_\beta}{m_\beta} \frac{1}{\DMS_{\alpha\beta}}\left(\vv_\beta-\vv_\alpha\right)
\end{align}
for $\alpha,\beta=a,b$. These equations are also valid for mixtures with more than two components. Notice the presence of the electrochemical potentials\index{electrochemical potential}
\begin{align}
\mue_\alpha = \mu_\alpha + \frac{z_a e_0}{m_a} \varphi,
\end{align}
where $z_a = q_a/e_0$ is the charge number and $e_0$ the elementary charge. Equations \eqref{eq.Maxtef} are referred to as the Maxwell-Stefan relations. \index{Maxwell-Stefan diffusion}
Equations \eqref{eq.EMHD2.MechEq.ra}, \eqref{eq.EMHD2.MechEq.rb}, \eqref{eq.Maxtef}, \eqref{eq.EMHD2.Poisson} are referred to as the generalized Poisson-Nernst-Planck equations, and they serve as the basic equations governing electrochemical processes. Note that, in addition, the total momentum vanishes, $0=\sum \rho_\alpha \uu_\alpha$, in mechanical equilibrium, which accompanies the above velocity relations. 

\subsection{Heat conduction}
We have seen that energy can be transported by a motion of matter and by the electromagnetic field. Another means of energy transport is the heat conduction. One can again formulate a detailed evolution equation for heat including the inertia of the heat flux \cite{dynmaxent}, but we shall restrict ourselves to the simplest possible description of heat flux, the Fourier law. The heat flux, which represents a part of energy flux, is proportional to the negative of the gradient of temperature so that heat flows to the lower temperatures, in accordance with the second law of thermodynamics,
\begin{equation}\label{eq.Fourier}\index{Fourier law}
\JJ_q = -\lambda \nabla T,
\end{equation}
where $\lambda$ is the coefficient of thermal conductivity. The balance of energy then becomes
\begin{equation}\label{eq.e.Fourier}
\partial_t e = -\nabla\cdot (J_e + J_q).
\end{equation}
The corresponding balance of entropy is
\begin{align}\label{eq.s.Fourier}
\partial_t s = -\nabla\cdot\left(\sum_\alpha s_\alpha \vv_\alpha + \frac{\JJ_q}{T}\right) + \JJ_q \cdot \nabla T^{-1}+ \frac{1}{2 T} \zeta_{ab}(\vv_a-\vv_b)^2,
\end{align}
so that Eq. \eqref{eq.e.Fourier} is satisfied. Note that the part of entropy balance that is not under the divergence is called entropy production,
\begin{equation}
\sigma_s = \frac{\lambda}{T^2}\nabla T \cdot \nabla T+ \frac{1}{4 T} \zeta_{ab}(\vv_a-\vv_b)^2,
\end{equation}
a clearly non-negative quantity.\index{entropy production}

The overall heat flux into the system is denoted by $Q$, \index{heat flux}
\begin{equation}
Q = -\int_{\partial \Omega} \JJ_q\cdot \dd S.
\end{equation}

\subsection{Example: free energy model of a regular binary solution}\label{sec.sa_example}
Let us consider a free energy model of a regular solution: a binary mixture of ideal gasses with two temperatures and a heat-of-solution term. The free energy, expressed in the number densities $n_\alpha = \frac{\rho_\alpha}{m_\alpha}$, reads
\begin{align}
  f(n_a, n_b, T_a, T_b)
  &= 
    \sum_{\alpha\in\{a,b\}}
  \underbrace{
    c_\alpha^\textrm{V} n_\alpha T_\alpha \left( 
      1-\ln \left(
          c_\alpha^\textrm{V} n_\alpha^{1-\gamma_\alpha} T_\alpha \Phi_\alpha
      \right)
    \right)
}_{
    \textrm{ideal gas}
}
  +\underbrace{
   (T_a \kappa_a + T_b \kappa_b)\frac{n_a n_b}{n_a+n_b}
  }_{
      \textrm{generalized regular solution}
  }
  ~,
  \label{eq.free-energy-reg-sol-2}
\end{align} 
where $c_\alpha^\textrm{V}$ is the isochoric heat capacity; $\gamma_\alpha$ is the adiabatic factor; $\Phi_\alpha$ is a constant that follows from statistical physics underpinning; and finally 
$\kappa_a$ and $\kappa_b$ are generalized heats of solution.

To illustrate the above theory, we identify the generalised Maxwell-Stefan governing equations. To this end, we identify the $\tilde{s}_a$ as we assume a detailed knowledge on the macroscale \eqref{eq.free-energy-reg-sol-2}. Although the partial entropies $s_\alpha$ have not been identified, the role of individual temperatures $T_\alpha$ is equivalent as $s_\alpha=-\frac{\partial f}{\partial T_\alpha}$.

The condition \eqref{eq.sab.T} defining the unknown functions $\tilde{s}_\alpha$ can be simplified in the current case when the total free energy is a known function of the partial temperatures as $\tilde{s}_\alpha=s_\alpha|_{T_\alpha=T}$. Hence, we have
\begin{equation}
  \label{eq:sa_reg-sol-2}
  \tilde{s}_\alpha = c_\alpha^\textrm{V} n_\alpha \ln \left(c_\alpha^\textrm{V} T n_\alpha^{1-\gamma_\alpha} \Phi_\alpha\right) - \frac{n_a n_b}{n_a+n_b} \kappa_\alpha.
\end{equation}

Further, the chemical potential (per kilogram, as defined above) reads
\begin{align*}
  \mu_\alpha =& \frac{\partial f}{\partial \rho_\alpha}\bigg|_{T_a=T_b=T}= \frac{\partial f\big|_{T_a=T_b=T}}{\partial \rho_\alpha}=\\
  &=\frac{T}{m_\alpha} \left[c_\alpha^\textrm{V} \gamma_\alpha+\frac{\kappa_a+\kappa_b}{(n_a+n_b)^2}n_\beta^2-c_\alpha^\textrm{V} \ln\left(c_\alpha^\textrm{V} T n_\alpha^{1-\gamma_\alpha} \Phi_\alpha\right)\right]
\end{align*}
where $\alpha,\beta\in\{a,b\}$ and $\alpha\neq\beta$.
\label{eq.CIT-regsol-potentials-2T}

The governing equations read
\begin{align*}
      \partial_t \rho_a   &= -\partial_i\left(\rho_a v_a^i \right)~,\\
       \partial_t \rho_b   &= \partial_i\left(\rho_a v_a^i \right)~,\\
  v_a^i  &= \frac{1}{\zeta_{ab}}\frac{\rho_b}{\rho}\left(  -\rho_a \partial_i \mu_a- \tilde{s}_a\partial_i T \right)\\
  =& \frac{1}{\zeta_{ab}}\frac{\rho_b}{\rho}\bigg( \frac{T}{(n_a+n_b)^3}\left[c_a^\textrm{V}(1-\gamma_a)n_a^3+3 c_a^\textrm{V}(1-\gamma_a) n_a^2 n_b+(3 c_a^\textrm{V}(1-\gamma_a)+2 (\kappa_a+\kappa_b)) n_a n_b^2+c_a^\textrm{V} (1-\gamma_a) n_b^3\right]\partial_i n_a\\
  &+ \frac{2 (\kappa_a+\kappa_b) T n_a^2 n_b}{(n_a+n_b)^3} \partial_i n_b\\
                          &+ \frac{n_a}{(n_a+n_b)^2}\left[c_a^\textrm{V}(1-\gamma_a) n_a^2+(2 c_a^\textrm{V}(1-\gamma_a)+\kappa_a) n_a n_b+(c_a^\textrm{V}(1-\gamma_a)-\kappa_b)n_b^2 \right]\partial_i T \bigg)\\
\partial_t s      &= -\partial_i\left[\left( \tilde s_a -\frac{m_a}{m_b} \tilde{s}_b\right)v_a^i\right]  + \frac{1}{4 T} \zeta_{ab}\left(1-\frac{m_a}{m_b}\right)^2\mathbf{v}_a^2.
    \end{align*}

To this end, we calculate the thermodynamic force of the component $\alpha$, 
which contributes to the momentum equations~\eqref{eq.EMHD2.CIT.u} and~\eqref{eq.EMHD2.CIT.w}, it reads
\begin{align}
  \nonumber  
  -n_\alpha\partial_i\mu_\alpha - \widetilde{s}_\alpha\partial_i T
  =& 
  -(\gamma_\alpha - 1)c_\alpha^\textrm{V} T \left(
    \partial_i n_\alpha 
    + n_\alpha \partial_i \ln T
  \right)
  - T(\kappa_a + \kappa_b ) 
  \frac{2 n_a n_b}{(n_a+n_b)^3}
  \left(
    n_\alpha\partial_i n_\beta
   -n_\beta\partial_i n_\alpha
  \right)
  \\
  &
  +(\kappa_a + \kappa_b ) \frac{n_a n_b^2 }{(n_a+n_b)^2} \partial_i T
  -\kappa_a  \frac{n_a n_b }{(n_a+n_b)}\partial_i T
 ~,
  \label{eq.CIT-reduced-driving-force-2T}
\end{align}
where $\alpha,\beta\in\{a,b\}$ and $\alpha\neq\beta$.
The coefficients before the temperature gradient $\partial_i T$ in this case read
\begin{align}
    -n_\alpha(\gamma_\alpha -1 )c_\alpha^\textrm{V} 
    +
    \underbrace{
      \frac{n_a n_b }{(n_a+n_b)^2}\left( 
      n_b\kappa_b   
      - n_a\kappa_a  
      \right)
    }_{
        \textrm{non-isothermal interaction}
    }
\end{align}
In the following Section, we show another application, namely the dusty gas model.

\section{Dusty gas limit of the Maxwell-Stefan equations}\label{sec.dgm}
Dusty gas is a widely applied model of porous medium, obtained as the limit of Maxwell-Stefan transport equations~\cite{mason},~\cite{krishna-wesselingh},~\cite{KerkhofGeboers05}.
To derive the dusty gas model, let us consider the system~\eqref{eq.EMHD2.MechEq} with an additional species, denoted with subscript $D$.
This additional species, representing the dust, is assumed to be static and uniformly-distributed. In other words, we assume that $n_D = const$ and $\vv_D = 0$.
Note that the last assumption is the main difference from the mechanical equilibrium case developed in Section~\ref{sec.mecheq}.
We adjust the diffusion coefficients, similarly as in~\eqref{eq.diffcoef} as in~\cite{mason}, we define $ \frac{\varepsilon}{\tau} n D_{\alpha\beta} = \DMS_{\alpha\beta}(n+n_D) $, $ \frac{\varepsilon}{\tau} D_{\alpha D} = \frac{n + n_D}{n_D} \DMS_{aD}$, where $\varepsilon$ is the dust's volume fraction and $\tau$ stands for the tortuosity.
The force-flux relation yields
\begin{align}\label{eq.DGM.force-flux}
     n_\alpha \partial_i \mol{\mu}_\alpha 
+ s_\alpha\partial_i T
+ z_\alpha e_0 n_\alpha\partial_i \varphi 
-
{k_B T}
\left(
    \sum_{\beta\neq D} \frac{1}{\frac{\varepsilon}{\tau}n D_{\alpha\beta}}\left(
        n_\alpha J^i_\beta - n_\beta J^i_\alpha
    \right)
    - \frac{1}{\frac{\varepsilon}{\tau}D_{\alpha D}} J^i_\alpha
\right)
&=
0
~,
\quad\mbox{for }\alpha\neq D
~,
\end{align}
where $n = \sum_{\alpha\neq D} n_\alpha$ and $J_\alpha^i = n_\alpha v_\alpha^i$.
The dusty gas limit of the Maxwell-Stefan system reads
\begin{subequations}
\begin{align}
    \partial_t n_\alpha + \partial_i J_\alpha^i 
    &= 0
    ~,
    \quad\mbox{for }\alpha\neq D
    ~,
   \\ 
    \partial_i \left( \varepsilon_0 \varepsilon_r \partial_i\varphi \right)
    - e_0 \sum_\alpha z_\alpha n_\alpha 
    &= 0
    ~,
    \\    
    \label{eq.DGM.system.energy}
    \partial_t \left( 
        \epsilon(n_\alpha, T) 
        + \frac{\varepsilon_0 \varepsilon_r}{2} \partial_i \varphi\partial_i \varphi
    \right) 
    +
    \partial_i\left( 
        -\lambda\partial_i T
        +\sum_{\alpha\neq D} \left( 
            \mu_\alpha + T\frac{\tilde{s}_\alpha}{n_a} 
        \right) J_\alpha^i
    \right) 
    &=0
    ~.
\end{align}
\label{eq.DGM.system}
\end{subequations}

\subsection{Ideal gas and dust -- Fick's law}
In particular for a mixture of neutral ideal gas and dust at constant temperature, the dusty gas limit of the Maxwell-Stefan equations \eqref{eq.DGM.force-flux} can be rewritten as
\begin{align}
	-\frac{\varepsilon}{\tau}D_{aD} \nabla n_a = \mathbf{J}_a,
\end{align}
which is the Fick's law, telling that flux of component $a$ is proportional to negative of the gradient of concentration of the component.\index{Fick's law}

\subsection{Nonisothermal binary mixture of neutral ideal gasses in a porous medium}
Let us illustrate the system~\eqref{eq.DGM.system} on an example of a mixture of two neutral ideal gasses and dust: $a$, $b$ and $D$. 
The free energy of the mixture reads
\begin{align}
\epsilon^*(n_a, n_b, T)
  &= c_a^V n_a T \left( 
      1-\ln \left(
        c_a^V n_a^{1-\gamma_a} T\Phi_a
      \right)
    \right)
   +c_b^V n_b T \left( 
          1-\ln \left(
            c_b^V n_b^{1 -\gamma_b} T\Phi_b
          \right)
    \right)
+c_D^V n_D T
  ~,
\label{eq.DGM2.free-energy}
\end{align}
where $c_\alpha^V$ is the isochoric heat capacity, $\gamma_\alpha$ is the adiabatic factor, and $\Phi_\alpha=e^{\gamma_\alpha} \frac{4\pi m_\alpha}{3 \hbar^2}$ is a constant that follows from statistical physics underpinning, cf.\ Sackur-Tetrode entropy. 
The reduced momentum equations read
\begin{subequations}
\begin{align}
    \frac{c_a^V}{k_B} (\gamma_a-1)\left(  
      \partial_i n_a + n_a \partial_i \ln T
     \right)
    &=
    \frac{n_a}{n_a + n_b}
    \frac{1 }{\frac{\varepsilon}{\tau}D_{ab}}
        J^i_b 
    -
    \left( 
          \frac{n_b }{n_a + n_b }\frac{1 }{\frac{\varepsilon}{\tau}D_{ab}}
          + \frac{1}{\frac{\varepsilon}{\tau}D_{aD}}
    \right)
    J^i_a
    ~,
    \\
    \frac{c_b^V}{k_B}  (\gamma_b-1)\left(  
    \partial_i n_b +n_b  \partial_i \ln T
    \right)
    &=
    \frac{n_b}{n_a + n_b }
    \frac{1}{\frac{\varepsilon}{\tau}D_{ab}}
        J^i_a 
    -
    \left(
        \frac{1}{\frac{\varepsilon}{\tau}D_{bD}} + \frac{n_a}{n_a + n_b }\frac{1}{\frac{\varepsilon}{\tau}D_{ab}}
    \right)
         J^i_b
    ~.
\end{align}
\label{eq.DGM2.FF}
\end{subequations}
A stationary numerical solution of the system~\eqref{eq.DGM.system} with Dirichlet boundary condition is illustrated in~\ref{fig:MOL}. The model formulation and the finite volume discretization are summarized in the ~\ref{sec.fvm}.

\subsubsection{Analysis of irreversibility of the nonisothermal binary mixture of neutral ideal gasses in a porous medium}
The particular form of the entropy production density follows from the general formula~\eqref{eq.DGM.entropy-production} and it reads
\begin{subequations}
\begin{align}
\sigma_\text{Fourier} 
=&\ 
\lambda\left(
	\frac{\partial_i T}{T}
\right)^2
\\
\sigma_\text{MS}
=&\
\frac{n_a n_b}{n_a+n_b}
\frac{k_B}{\frac{\varepsilon}{\tau}D_{ab}}
\left(
	\frac{J_a^i}{n_a}
	-
	\frac{J_b^i}{n_b}
\right)^2
\\
\sigma_\text{dust}
=&\
{n_a}
\frac{k_B}{\frac{\varepsilon}{\tau}D_{aD}}
\left(
	\frac{J_a^i}{n_a}
\right)^2
+
{n_b}
\frac{k_B}{\frac{\varepsilon}{\tau}D_{bD}}
\left(
	\frac{J_b^i}{n_b}
\right)^2
\\
\sigma 
=&\ \sigma_\text{Fourier} + \sigma_\text{MS} + \sigma_\text{dust}
\label{eq.DGM.entropy-production}
\end{align}
\end{subequations}

\begin{figure}
    \centering
  \includegraphics[width=0.6\textwidth]{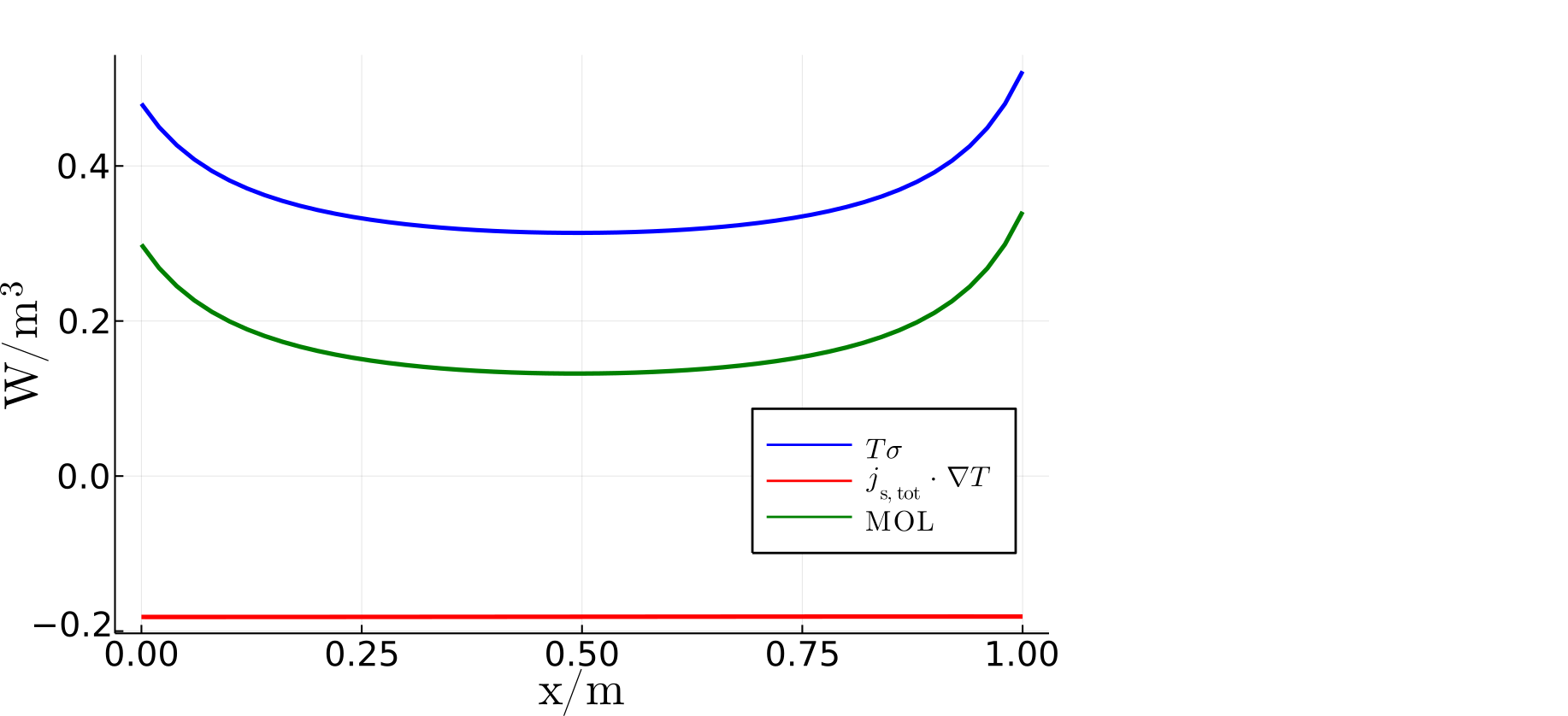}
  \label{fig:MOL}
  \caption{Map of losses (MOL), see~\eqref{eq:MOL} shows the density of energy losses due to the binary diffusion and heat conduction in a porous medium~\eqref{eq:num_dgm}.}
\end{figure}

We can now plot a quantity assessing the locations of lost power. Once we know what is a useful power in a given problem, typically determined by what can be controlled in a given system, we can define the map of losses $\textrm{MOL}(x)$--the density of lost power\footnote{energy dissipation rate}. Let us consider that we can control the mass fluxes across the boundary but not temperature and temperature gradients. Then, the map of losses can be retrieved from the stationary energy balance~\eqref{eq.DGM.system.energy} as follows
\begin{align}
0 =&
\int_\Omega
\partial_i\left(
	-\lambda\partial_i T
	+T\frac{\tilde s_a}{n_a}J^i_a
	+T\frac{\tilde s_b}{n_b}J^i_b
\right)
+
\partial_i\left(
	\mu_a J^i_a
	+\mu_b J^i_b
\right)
\nonumber
\\
=& 
\int_\Omega
\partial_i\left(
	T\left[
		-\lambda\frac{\partial_i T}{T}
		+\frac{\tilde s_a}{n_a}J^i_a
		+\frac{\tilde s_b}{n_b}J^i_b
	\right]
\right)
+
\partial_i\left(
	\mu_a J^i_a
	+\mu_b J^i_b
\right)
\nonumber
\\
=& 
\underbrace{
	\int_{\partial\Omega}
	\left(
		\mu_a J^i_a
		+\mu_b J^i_b
	\right)
	\cdot \nu_i
}_{\dot\Delta G}
+
\int_\Omega
\underbrace{
	\left[
	\partial_i
		T\left(
			-\lambda\frac{\partial_i T}{T}
			+\frac{\tilde s_a}{n_a}J^i_a
			+\frac{\tilde s_b}{n_b}J^i_b
		\right)	
		+
		T\sigma
	\right]
}_{=:MOL(x)}
\label{eq:MOL}
~.
\end{align}
The spatial profile of $\textrm{MOL}(x)$ is shown in Fig.~\ref{fig:MOL}, the entropy production density overestimates the power losses which supports the findings in~\cite{pavelka-ae} and~\cite{vagner2017pitfalls}.

\section{Full reduction to CIT}\label{sec.full}
The reductions described in the preceding two sections are only approximate, as the reduction is not complete, the reduced equations are not closed with respect to the lower state variables (e.g. $s_\alpha$ still persist). One requires to provide some information from the detailed level, detailed entropy in our case, to be able to calculate $\tilde{s}_a,~\tilde{s}_b$. 

Here, we take a different route where the reduction is complete and the reduced equations are closed with respect to the less detailed state variables. The price we pay is the complexity of the obtained equations. To this end, we consider a reduction in direct and conjugate state variables. As we shall see, the reduction in conjugate variables is more natural and general because it does not require the particular knowledge of the detailed entropy. In this sense, the approach presented above and the projection to CIT in direct variables is similar as they both require the specification of the detailed entropy to proceed.

\subsection{Projection to CIT in direct variables}
Starting point is the binary mixture description without charge \cite{PKG}, i.e. Eqns \eqref{eq.EMHD2} where $\EE=\BB=0$. With an explicit choice of entropy (entropy of ideal gases) we may use the dynamic MaxEnt \cite{PKG,RedExt} to identify the  evolution equations on the CIT level in the mechanical equilibrium, i.e. state variables $\rho_a,~ \rho_b,~ \uu=\uu_a+\uu_b,~e=e_a+e_b$ where momentum $\uu$ is constant.

The dynamics on the microscale corresponding to binary fluid mechanics $\rho_a,~\rho_b,~ \uu_a,~\uu_b,~s_a,~s_b$ is given by 
\begin{subequations}
\begin{equation}
    \frac{\partial \rho_\alpha}{\partial t} = -\div(\rho_\alpha \uu_\alpha^\dagger),
\end{equation}
\begin{equation}
    \frac{\partial u_{\alpha,i}}{\partial t} = -\rho_\alpha \partial_i \rho_\alpha^\dagger - u_{\alpha,j} \partial_i u_{\alpha,j}^\dagger - s_\alpha \partial_i s_\alpha^\dagger - \partial_j (u_{\alpha,i} u_{\alpha,j}^\dagger)-(\delta_{a\alpha}-\delta_{b \alpha})\zeta_{ab}(u_{a,i}^\dagger-u_{b,i}^\dagger)
\end{equation}
\begin{equation}
  \frac{\partial s_\alpha}{\partial t} = -\div(s_\alpha \uu_\alpha^\dagger)+\frac{1}{2}\frac{s_\alpha^\dagger}{(s_a^\dagger+s_b^\dagger)^2}\sum_{\beta,\gamma}\zeta_{\gamma\beta}(\uu_\gamma^\dagger-\uu_\beta^\dagger)^2-\frac{2}{s_a^\dagger+s_b^\dagger} K_{ab}(s_a^\dagger-s_b^\dagger)(\delta_{b\alpha}s_a^\dagger-\delta_{a \alpha}s_b^\dagger),
\end{equation}
\end{subequations}
with a typical energy
\begin{equation}
  E = \int \mathrm{d} \rr \left( \frac{\uu_a^2}{2 \rho_a} + \frac{\uu_b^2}{2 \rho_b} + \epsilon(\rho_a,s_a,\rho_b,s_b)\right).
\end{equation}
The effect of temperature equilibration and mutual friction is described via a dissipation potential as above, eq. \eqref{eq.Xi} while, to complete the description, we consider the entropy potential for a binary mixture of ideal gases
\begin{equation}
  S(\rho_a,\rho_b,\uu_a,\uu_b,e_a,e_b) = S_a+S_b =  \sum_{\alpha\in\{a,b\}}\int \mathrm{d}\rr k_B \frac{\rho_\alpha}{m_\alpha}\left(\frac{5}{2} + \ln \left[\frac{m_\alpha}{\rho_\alpha} \left(\frac{4 \pi m_\alpha}{3 h^2} \frac{e_\alpha-\frac{\uu_\alpha^2}{2 \rho_\alpha}}{\rho_\alpha/m_\alpha}\right)^{3/2}\right]\right).
\end{equation}

Finally, the classical irreversible thermodynamics (CIT) level is characterised by the following choice of state variables $\rho_a,~\rho_b,~\uu,~s=s_a+s_b$. It is again advantageous to change the state variables in binary hydrodynamics prior to this projection to CIT level so that one can easily track the effects of different temperatures or dissipation related to the relative motion. To this end we again introduce $$\uu = \uu_a+\uu_b, ~\mathbf{w} = 1/2(\uu_a-\uu_b)$$ and define $$s=s_a+s_b, ~\sigma = 1/2(s_a-s_b).$$ Similarly as for $u^\dagger,~w^\dagger$ discussed above, one can show \cite{PKG} that the conjugate variables are related via $$\uu^\dagger = 1/2(\uu_a^\dagger+\uu_b^\dagger),~\mathbf{w}^\dagger = \uu_a^\dagger-\uu_b^\dagger, ~ s^\dagger=1/2(s_a^\dagger+s_b^\dagger),~ \sigma^\dagger = s_a^\dagger-s_b^\dagger.$$

With the above explicit choice of entropy, the MaxEnt (MaxEnt values of state variables are denoted with ME in the upper index) gives
\begin{equation*}
  \frac{\uu_\alpha^{ME}}{\rho_\alpha} = \frac{\uu}{\rho_T},\quad \rho_T= \rho_a+\rho_b,
\end{equation*}
hence the most probable way of distributing the total momentum among partial momenta is via mass fractions. Similarly MaxEnt values of partial energies satisfy
\begin{equation*}
  \frac{e_\alpha^{ME}-(\uu_\alpha^{ME})^2/(2\rho_\alpha)}{\rhob_\alpha} = \frac{e - \uu^2/(2 \rho_T)}{\rhob_T},
\end{equation*}
where $\rhob_\alpha=\rho_\alpha/m_\alpha$.

The explicit knowledge of entropy can be used to identify $\uu=0$ in the mechanical equilibrium and similarly $\mathbf{w}^{ME} = \frac{1}{2}(\uu_a-\uu_b)= \frac{\mathbf{u}}{2} \frac{\rho_a-\rho_b}{\rho_a+\rho_b}=0$ while
\begin{equation} \label{eq.sigmaME}
  \sigma^{ME}(\rho_a,\rho_b,s)=\frac{1}{2}(s_a-s_b) = -\frac{5}{2} k_B \frac{\rhob_a-\rhob_b}{2} + 2k_B \frac{\rhob_a\rhob_b}{\rhob_T} \ln \left(\left(\frac{m_a}{m_b}\right)^{3/2}\frac{\rhob_b}{\rhob_a}\right) + \frac{\rhob_a-\rhob_b}{\rhob_T} s,
\end{equation}
where we introduced $\rhob_T = \rhob_a+\rhob_b$.

Dynamic MaxEnt provides a relation for direct state variables of the relaxed state variables from the higher, more detailed, level of description (in our case $\mathbf{w}, \sigma$). The values of their conjugate variables have to be identified so that the dynamics of the relaxed evolution on the lower level stays exactly on the MaxEnt values in the direct variables.

Hence $\mathbf{w}_\alpha^\dagger$ and $\sigma^\dagger$ have to be identified from their evolution equations following from their Max Ent values. In particular,
\begin{subequations} \label{Eq.4ualphaDaggerME}
\begin{equation} \label{Eq.4uaDaggerME}
  \partial_t \mathbf{u}_a^{ME}  = \boxed{ 0 = -\rho_a \nabla \rho_a^\dagger - \zeta_{ab} (\uu_a^\dagger - \uu_b^\dagger)} - \underbrace{(s/2+\sigma^{ME})}_{s_a}\nabla \underbrace{(s^\dagger+\sigma^\dagger/2)}_{s_a^\dagger},
  \end{equation}
  \begin{equation} \label{Eq.4ubDaggerME}
   \partial_t \mathbf{u}_b^{ME} = \boxed{ 0 =  -\rho_b \nabla \rho_b^\dagger - \zeta_{ab} (\uu_b^\dagger - \uu_a^\dagger)} - \underbrace{(s/2-\sigma^{ME})}_{s_b}\nabla \underbrace{(s^\dagger-\sigma^\dagger/2)}_{s_b^\dagger},
 \end{equation}
 \end{subequations}
 where the boxed terms correspond to the classical Maxwell-Stefan model. The MaxEnt value of $\sigma$, $\sigma^{ME}$, can be used to get
 \begin{subequations} \label{Eq.4sigmaDaggerME}
\begin{align}
  \partial_t   \sigma^{ME} &= -\frac{1}{2} \div\left[(\uu_a^\dagger+\uu_b^\dagger)\left(-\frac{5}{2} k_B \frac{\rhob_a-\rhob_b}{2} + 2 k_B \frac{\rhob_a \rhob_b}{\rhob_T} \ln\left[\left(\frac{m_a}{m_b}\right)^{3/2} \frac{\rhob_b}{\rhob_a}\right]\right)\right] \\
  &- \frac{1}{4} \div\left[ \frac{s}{\rhob_T} \left(\uu_a^\dagger(3\rhob_a-\rhob_b) - \uu_b^\dagger (3 \rhob_b-\rhob_a)\right)\right] - K_{ab} \sigma^\dagger\nonumber\\
                           &=\partial_t \rhob_a \left[-\frac{5}{4} k_B + 2 k_B \frac{\rhob_b}{\rhob_T}\ln\left[\left(\frac{m_a}{m_b}\right)^{3/2} \frac{\rhob_b}{\rhob_a}\right] - 2k_B \frac{\rhob_b}{\rhob_T} + 2\frac{\rhob_a}{\rhob_T} s \right]\\
                             &+\partial_t \rhob_b \left[\frac{5}{4} k_B + 2 k_B \frac{\rhob_a}{\rhob_T}\ln\left[\left(\frac{m_a}{m_b}\right)^{3/2} \frac{\rhob_b}{\rhob_a}\right] + 2k_B \frac{\rhob_a}{\rhob_T} - 2\frac{\rhob_b}{\rhob_T} s \right]+\partial_t s \frac{\rhob_a-\rhob_b}{\rhob_T},\nonumber
\end{align}
 \end{subequations}
where the first equality follows from the general evolution equations for $\sigma$, i.e. $\partial_t \sigma = \frac{1}{2}(\partial_t s_a - \partial_t s_b)$, with $s_a=\frac{s}{2}+\sigma^{ME}$, $s_b=\frac{s}{2}-\sigma^{ME}$, while the second equality comes from the time derivative of the MaxEnt estimate of $\sigma$, $\sigma^{ME}$ in Eq. \eqref{eq.sigmaME}.

These three relations, Eqs \eqref{Eq.4ualphaDaggerME}, \eqref{Eq.4sigmaDaggerME}, for the conjugate values of the relaxed state variables, $\uu_a^\dagger,~\uu_b^\dagger,~\sigma^\dagger$, provide closures for the evolution of the lower state variables
\begin{align}
  \label{Eq.EvoLower}
  \partial_t \rho_\alpha &= - \div (\rho_\alpha \uu_\alpha^\dagger),\\
  \partial_t s &= - \div\left\{(\uu_a^\dagger-\uu_b^\dagger)\left(-\frac{5}{2} k_B \frac{\rhob_a-\rhob_b}{2} + 2 k_B \frac{\rhob_a \rhob_b}{\rhob_T} \ln\left[\left(\frac{m_a}{m_b}\right)^{3/2} \frac{\rhob_b}{\rhob_a}\right]\right) + \frac{1}{2} \frac{s}{\rhob_T} \left(\uu_a^\dagger(3\rhob_a-\rhob_b) + \uu_b^\dagger (3 \rhob_b-\rhob_a)\right)\right\} +\nonumber\\ &+\frac{1}{s^\dagger} \left[\frac{1}{4}\zeta_{ab}(\uu_a^\dagger - \uu_b^\dagger)^2 + K_{ab} \left(\sigma^\dagger\right)^2\right].
\end{align}

The standard Maxwell-Stefan model is obtained (the difference in the conjugate momenta provide a relation for the chemical potentials $\rho_\alpha^\dagger$) when the entropy is allowed to equilibrate (when the temperature, $s^\dagger$, is such that $\partial_t s=0$). In the extended version with the temperature effect, one has to solve for $\sigma^\dagger$ which decouples from the $\uu_a^\dagger,~\uu_b^\dagger$ dependency once the two relations \eqref{Eq.4ualphaDaggerME} are added.

\subsubsection{Fast dissipation}
In the case of a fast dissipation (the time scale required for the entropy and momentum dissipation is significantly shorter than the time scale of reversible evolution), we may assume that the terms $K_{ab}$ and $\zeta_{ab}$ are significantly larger than the remaining ones (as in Chapman-Enskog solution of Boltzmann equation). Then the leading order equation for $\sigma^\dagger$ reads from \eqref{Eq.4sigmaDaggerME} as
\begin{equation*}
  0=K_{ab} \sigma^\dagger + \frac{1}{s^\dagger} K_{ab} \left(\sigma^\dagger\right)^2 \frac{\rhob_a-\rhob_b}{\rhob_T},
\end{equation*}
with two leading order solutions
\begin{align}
  \label{Eq.sigmaDaggerLeadingOrder}
  \sigma^\dagger = 0,\\
  \sigma^\dagger = -s^\dagger \frac{\rhob_T}{\rhob_a-\rhob_b}.
\end{align}
The second solution, however, is not a physically admissible as one of the corresponding partial temperatures would have to be negative
\begin{equation*}
  s_a^\dagger = s^\dagger+\sigma^\dagger/2 = -s^\dagger\frac{2 \rhob_b}{\rhob_a-\rhob_b},\quad   s_b^\dagger = s^\dagger-\sigma^\dagger/2 = s^\dagger\frac{2 \rhob_a}{\rhob_a-\rhob_b}.
\end{equation*}
However, the first leading order solution is always plausible and natural as it corresponds to the same partial temperatures in CIT mixture where one measures just a single temperature without the possibility to distinguish momenta of each constituent.

In this setting the extended version of Maxwell-Stefan model simplifies to 
\begin{align}
  \label{Eq.EvoLowerLeadingOrder}
 &\boxed{\partial_t \rho_\alpha = - \div (\rho_\alpha \uu_\alpha^\dagger)},\\
 &\boxed{ 0 = -\rho_a \nabla \rho_a^\dagger - \zeta_{ab} (\uu_a^\dagger - \uu_b^\dagger)} - (s/2+\sigma^{ME}(\rho_a,\rho_b,s))\nabla s^\dagger,\\
   &\boxed{ 0 =  -\rho_b \nabla \rho_b^\dagger - \zeta_{ab} (\uu_b^\dagger - \uu_a^\dagger)} - (s/2-\sigma^{ME}(\rho_a,\rho_b,s))\nabla s^\dagger,\\
&  \partial_t s = - \div\left\{(\uu_a^\dagger-\uu_b^\dagger)\left(-\frac{5}{2} k_B \frac{\rhob_a-\rhob_b}{2} + 2 k_B \frac{\rhob_a \rhob_b}{\rhob_T} \ln\left[\left(\frac{m_a}{m_b}\right)^{3/2} \frac{\rhob_b}{\rhob_a}\right]\right) + \frac{1}{2} \frac{s}{\rhob_T} \left(\uu_a^\dagger(3\rhob_a-\rhob_b) + \uu_b^\dagger (3 \rhob_b-\rhob_a)\right)\right\} +\\ &+\frac{1}{s^\dagger} \zeta_{ab}(\uu_a^\dagger - \uu_b^\dagger)^2.
\end{align}
Note again that, in addition, the total momentum vanishes, $0=\sum \rho_\alpha \uu_\alpha$, in mechanical equilibrium, which closes the problem.

Finally note that $s^\dagger =T$ and using the assumption of the explicit form of entropy one can rewrite both $s$ and $s^\dagger$ in terms temperature ($e=\frac{3}{2} k_B \rhob_T T$ in MaxEnt):
\begin{equation*}
  s = \frac{5}{2}\rhob_T k_B + k_B \left[\rhob_a \ln\left(\frac{m_a^{3/2}}{\rhob_a}\right)+\rhob_b \ln\left(\frac{m_b^{3/2}}{\rhob_b}\right)\right]+k_B \rhob_T \frac{3}{2} \ln\left(\frac{ 2 \pi}{h^2} k_B T\right).
\end{equation*}

It is a straightforward calculation to show that total energy is conserved via
\begin{equation*}
  \partial_t e =  s^\dagger \partial_t s +  \rho_a^\dagger \partial_t \rho_a +  \rho_b^\dagger \partial_t \rho_b,
\end{equation*}
as energy change occurs only via fluxes.

%

\subsection{Projection to CIT in conjugate variables}

An alternative method of extension is via a projection from the following binary mixture state variables $\rho_a,~\rho_b,~\uu_a,~\uu_b,~s_a^\dagger,~s_b^\dagger$ to the lower level $\rho_a,~\rho_b,~s$ where the projection is such that temperature differences disappear, $\sigma^\dagger =0 $, while reaching again mechanical equilibrium.

This approach has the advantage of generality (it is not necessary to specify entropy relation or look for the leading order solutions) however, as we shall see, the resulting equations are somewhat more complex.

First, we need to rewrite the description of binary mixture to the new choice of state variables. For this we use
\begin{equation*}
  E^* = -E + s_a^\dagger s_a + s_b^\dagger s_b,
\end{equation*}
a free-energy. If we denote $\rho_a^\ddagger = \left(\frac{\partial e^*}{\partial \rho_a}\right)_{\rho_b,\uu_a,\uu_b,s_a^\dagger,s_b^\dagger}$ then
\begin{align*}
  \rho_\alpha^\ddagger &= -\rho_\alpha^\dagger,\\
  \uu_\alpha^\ddagger &= -\uu_\alpha^\dagger,\\
  \left(s_\alpha^\dagger\right)^\ddagger & = s_\alpha.
\end{align*}

The evolution equations for conjugate variables are
\begin{align*}
  \partial_t s_a^\dagger &= \frac{1}{\det \Hess(e^*)} \Bigg[\frac{\partial^2 e^*}{\partial (s_b^\dagger)^2}\underbrace{\left(\partial_t s_a - \frac{\partial^2 e^*}{\partial s_a^\dagger \rho_a} \partial_t \rho_a- \frac{\partial^2 e^^*}{\partial s_a^\dagger \rho_b} \partial_t \rho_b- \frac{\partial^2 e^^*}{\partial s_a^\dagger \uu_a} \partial_t \uu_a- \frac{\partial^2 e^^*}{\partial s_a^\dagger \uu_b} \partial_t \uu_b\right)}_{\boxed{1}}-\\
                           &-\frac{\partial^2 e^*}{\partial s_a^\dagger \partial s_b^\dagger}\underbrace{\left(\partial_t s_b - \frac{\partial^2 e^^*}{\partial s_b^\dagger \rho_a} \partial_t \rho_a- \frac{\partial^2 e^^*}{\partial s_b^\dagger \rho_b} \partial_t \rho_b- \frac{\partial^2 e^^*}{\partial s_b^\dagger \uu_a} \partial_t \uu_a- \frac{\partial^2 e^^*}{\partial s_b^\dagger \uu_b} \partial_t \uu_b\right)}_{\boxed{2}}\Bigg],\\
  \partial_t s_b^\dagger &= \frac{1}{\det \Hess(e^*)} \left[\frac{\partial^2 e^*}{\partial (s_b^\dagger)^2} \boxed{2} - \frac{\partial^2 e^*}{\partial (s_a^\dagger)^2} \boxed{1}\right].
\end{align*}

Instead of using MaxEnt we prescribe the projection as equal partial temperatures on the lower lever, i.e. $\sigma^\dagger = T_a-T_b=0$ while $s^\dagger = \frac{1}{2}(s^\dagger_a+s^\dagger_b)$.


With the assumption of mechanical equilibrium, the lower level is given by $\sigma^\dagger =0, \uu_a=\uu_b=0$ (i.e. no dissipation due to the mutual friction or temperature equilibration) while the conjugate $(\sigma^\dagger)^\ddagger=\sigma$ variable is given by
\begin{subequations} \label{Eq.conjugateMS}
\begin{align}
  \label{Eq.sigmaMaxEnt}
  0=\partial_t \sigma^\dagger = \div\left(\left(\frac{s}{2}+\sigma\right)\uu_a^\ddagger\right)-\div\left(\left(\frac{s}{2}-\sigma\right)\uu_b^\ddagger\right) + \div(\rho_a \uu_a^\ddagger)\left(\frac{\partial^2 e^*}{\partial s_b^\dagger \partial \rho_a}-\frac{\partial^2 e^*}{\partial s_a^\dagger \partial \rho_a}\right) + \div(\rho_b \uu_b^\ddagger)\left(\frac{\partial^2 e^*}{\partial s_b^\dagger \partial \rho_b}-\frac{\partial^2 e^*}{\partial s_a^\dagger \partial \rho_b}\right).
\end{align}

This constitutive relation can then be used to close the extended Maxwell-Stefan model
\begin{align}
  \label{Eq.MSconjugate}
   &\boxed{\partial_t \rho_\alpha =  \div (\rho_\alpha \uu_\alpha^\ddagger)},\\
 &\boxed{ 0 = \rho_a \nabla \rho_a^\ddagger + \zeta_{ab} (\uu_a^\ddagger - \uu_b^\ddagger)} - (s/2+\sigma)\nabla s^\dagger,\\
   &\boxed{ 0 =  \rho_b \nabla \rho_b^\ddagger + \zeta_{ab} (\uu_b^\ddagger - \uu_a^\ddagger)} - (s/2-\sigma)\nabla s^\dagger,\\
   &  \partial_t s^\dagger = \frac{1}{\det \Hess(e^*)} \left(\frac{\partial^2 e^*}{\partial (s_b^\dagger)^2}-\frac{\partial^2 e^*}{\partial (s_a^\dagger)^2}\right)
     \left[\div\left(\left(\frac{s}{2}+\sigma\right)\uu_a^\ddagger\right) - \div(\rho_a\uu_a^\ddagger)\frac{\partial^2 e^*}{\partial s_a^\dagger \partial \rho_a} -\div(\rho_b\uu_b^\ddagger) \frac{\partial^2 e^*}{\partial s_a^\dagger \partial \rho_b}\right],
\end{align}
\end{subequations}
where $s^\dagger = T$ is temperature and $s$ is related to it via the free energy $e^*$ as $s=(s^\dagger)^\ddagger$.

\section{Soret coefficient as appearing from the multiscale analysis}\label{sec.soret}

An important outcome of our multiscale Maxwell-Stefan models is the ability to assess the Soret coefficient based on a detailed description. The Soret coefficient is a coefficient between the density or composition gradient and the temperature gradient in a steady state characterised by a zero mass flux
\begin{equation}
  \label{eq:SoretCoef}
  \nabla \rho_\alpha = -L_{\rho_\alpha T} \nabla T,
\end{equation}
where $L_{\rho_\alpha T}$ is the Soret coefficient describing the thermal diffusion of component $\alpha$ (positive Soret coefficient have such components which concentrate in colder regions).

Although the Soret coefficient has been studied for a long time in non-equilibrium thermodynamics (being an off-diagonal entry of the Onsager matrix relating thermodynamic fluxes and forces \cite{dgm}), it is much less explored than for instance the diffusion coefficient. A comprehensive review can be found in \cite{kjelstrup}. The Soret coefficient can be approximated by means of kinetic theory \cite{hirschfelder,tichacek}, where it comes from the collision integral or activation energies, or by using non-equilibrium statistical mechanics \cite{kempers1989}. Finally, a comprehensive comparison between the theoretical results and experiments can be found in \cite{kempers2001}, where two origins of Soret coefficient are identified: the kinetic origin (different properties of collisions of the constituents), and the thermodynamic origin (different attraction/repulsion between the constituents). In this work, we show another possible origin -- different behavior in the two-temperature state (before relaxation of temperatures of the constituents to their common value).

In our study, the flux-force relations provide (noting that the flux vanish as long as one of the components disappears)
\begin{equation*}
  \rho_a v_a^i = -\frac{\rho_a \rho_b}{\zeta_{ab} \rho} \left(\rho_a \partial_i \mu_a+s_a \partial_i T\right)=0,
\end{equation*}
where the expression for $s_a$ depends on the chosen reduction. In the first approximate reduction where we assume detailed knowledge of entropy, $s_a$ is given by $\tilde{s}_a$ while in the full reductions they follow from MaxEnt estimate either in the direct or conjugate state variables.

By splitting the chemical potential gradient into the temperature gradient contribution and chemical potential potential of species $a$ at a constant temperature, we have
  \begin{equation*}
    (\partial_i \mu_a)_T = -\left[\frac{\partial \mu_a}{\partial T} +\frac{s_a}{\rho_a}\right]\partial_i T.
  \end{equation*}

  It is instructive to compare the various reductions. First, let us consider the approximative reduction in the first part of this paper, i.e. $s_a=\tilde{s}_a = s_a\big|_{T_a=T}$.
  Hence, Soret coefficient $L_{\rho_\alpha T}$ satisfies
  \begin{equation}
    \label{eq:Soret}
    L_{\rho_a T} = \frac{\frac{\partial \mu_\beta}{\partial T}+\frac{\tilde{s}_\beta}{\rho_\beta}}{\frac{\partial \mu_\beta}{\partial \rho_a}}=\frac{1}{\frac{\partial \mu_\beta}{\partial \rho_a}} \left[\frac{\partial \mu_\beta}{\partial T}-\frac{\partial \mu_\beta}{\partial T_\beta}\bigg|_{T_\beta=T}+\frac{1}{\rho_\beta}(\tilde{s}_\beta-s_\beta\big|_{T_\beta=T})\right]=\frac{1}{\frac{\partial \mu_\beta}{\partial \rho_a}} \left[\frac{\partial \mu_\beta}{\partial T}-\frac{\partial \mu_\beta}{\partial T_\beta}\bigg|_{T_\beta=T}\right].
  \end{equation}
  Hence, a nontrivial Soret coefficient is appearing only as an effect of a nonzero difference between partial temperatures.  If we consider the particular example described above in Section \ref{sec.sa_example}, we obtain
  \begin{equation*}
    L_{\rho_a T} = \frac{m_b}{ n_a (n_a+n_b)^2 T} \left(n_b\frac{(n_a+n_b)^3 \kappa_b}{\kappa_a+\kappa_b}-\frac{2 \kappa_a n_a^3}{\frac{(1-\gamma_b) c_b^\textrm{V}}{n_b}-\frac{2 n_a^2 (\kappa_a+\kappa_b)}{(n_a+n_b)^3}}\right).
  \end{equation*}

On the other hand, if we use the results of the full reduction, we obtain direct expressions for $s_a$ in terms of the less detailed state variables without a clear link to the necessity of having distinct partial temperatures. Hence, our study suggests that a temperature difference on a microscale manifests on a macroscale as a Soret effect of thermodiffusion, which provides an alternative origin of the Soret coefficient.

\section{Discussion and Conclusion}

Transport models are ubiquitous, attract a lot of attention, and are widely applicable, see for example the recent monograph \cite{venerus2018modern}. Here we focused on the extension of the successful diffusion Maxwell-Stefan model to the nonisothermal case to obtain a minimalistic and yet plausible description of thermodiffusion. To this end, we consider momentum exchange among the mixture components and keep track of the total entropy. To do so, we adapt a thermodynamic multiscale framework and employ reduction techniques to reach the desired level of detail.

In order to derive the hierarchy of theories of mixtures with two momenta and two temperatures, one momentum and one temperature, and finally without any momenta (mechanical equilibrium), we start from the Liouville equation and reduce this complex behavior by Hamiltonian reductions. This way, we obtain a hierarchy of models for multiscale binary homogeneous mixtures interacting with electrodynamics. In particular we obtain new terms in the equations for binary mixtures that can not be obtained by the standard approach based on balance equations.

The first approach, Sections \ref{sec.gen}-\ref{sec.dgm}, is the most intuitive one and easiest to use. It extends the classical Maxwell-Stefan to the non-isothermal case and, in addition, includes the effect of electric field via electro-chemical potential. Moreover it is straightforwardly extendable to a multicomponent mixture. However, the drawback is that in this intuitive approach we circumvented the problem of reducing the number of state variables entailing the presence of partial entropies in the final set of equations. Hence the problem is not closed and some form of further approximation have to take place.

The second approach, Section \ref{sec.full}, invoking MaxEnt to properly carry out the reduction to the lower level is more precise and yields a closed system of evolution equations. However, in order to explicitly carry out the reduction we needed to employ a particular choice of the upper entropy. If the ideal gas mixture is not appropriate, one has to repeat the reduction. Additionally, the temperatures of the two constituents are subdued to the lower state variables and are, in general, not equal. Note that when dissipation is dominant in the lower level evolution, one can show that the difference in temperatures vanishes.

Finally, the last approach used again MaxEnt dynamic reduction to yield the lower level evolution being the extension of Maxwell-Stefan equations. This time, the MaxEnt is carried out in the conjugate rather than direct state variables as to control the temperature difference of the two constituents being zero. Additionally, the yielded evolution equations are general -- one can chose free energy as the last step and thus specify the evolution equations.

Note that in principle all these approaches are extendable to a mixtures with a higher number of constituents but in the latter two approaches we do not include such extensions explicitly here. In addition, our approach reminds, at least in principle, the model of Lorentz diffusive gas \cite{AlvarezJou2007LorentzGas}, where the need to include momentum exchange among constituents resulted in evolution equations for new state variables in the spirit of Extended irreversible thermodynamics. The obtained governing equations are also Maxwell-Stefan-like equations. 

We apply the obtained Maxwell-Stefan theory to a dusty gas model, which we numerically solve to illustrate the novelty of the nonisothermal extension of the theory. One can see such effects directly from the equations themselves as, for example from \eqref{Eq.conjugateMS} one can expect nonfickian behaviour even in binary mixtures due to the presence of temperature gradients. Another application of our models is a non-isothermal generalization of the dusty gas model, its generalized thermodynamic efficiency analysis, and finally the identification of a new origin of the Soret coefficient. 

In future, we would like test the obtained equation in fast processes where the inertial effects and new terms might play a significant role.



\section*{Acknowledgments}
	MP was supported by project No. UNCE/SCI/023 of Charles University Research program. VK and MP were also supported by the Czech Science Foundation (project no. 20-22092S).
        PV was supported by the Deutsche Forschungsgemeinschaft (DFG, German Research Foundation) under Germany’s Excellence Strategy – The Berlin Mathematics Research Center MATH+ (EXC-2046/1, project ID: 390685689).
\bibliographystyle{plain}
\bibliography{library}

\appendix
\section{MaxEnt is equivalent to MinEne}
Let us now recall that maximization of concave entropy $S(E,x)$ with respect to $x$ is equivalent to minimization of energy $E(S,x)$, as shown for instance in \cite{callen,gibbscw}.

Assume a stationary MaxEnt point
\begin{equation}
    \left(\frac{\partial S}{\partial x}\right)_E\Big|_{\bar{x},E} = 0.
\end{equation}
By the implicit function theorem one gets
\begin{equation}
    \left(\frac{\partial E}{\partial x}\right)_S =
    -\frac{\left(\frac{\partial S}{\partial x}\right)_E}{\left(\frac{\partial S}{\partial E}\right)_x} = 0
\end{equation}
and thus 
\begin{equation}
    \left(\frac{\partial E}{\partial x}\right)_S\Big|_{\bar{x},S(E,\bar{x})} = 0.
\end{equation}
Energy thus also has a stationary point at $\bar{x}$.

Positivity of temperature means that entropy grows as energy grows (keeping $x$ constant). 

Imagine now a the surface $\sigma$ described by $S(x,E)=\const$ within the three-dimensional space $(E,x,S)$. Plane $\sigma_E$ is a plane satisfying $E=\const$. The point with $\bar{x}$ is the point with the highest value $S$ from the curve $\gamma_E = \sigma \cap \sigma_E$. Plane $\sigma_S$ is the plane characterized by $S=\const$ containing the point $\bar{x}$. Because entropy grows with energy, the curve $\gamma_S = \sigma_S\cap \sigma$ has its point with lowest energy located at $\bar{x}$ while other points of that curve have higher energy. Curves $\gamma_E$ and $\gamma_S$ have only one common point, namely $\bar{x}$. Maximization of entropy along the curve $\gamma_E$ is equivalent to miminization of energy along the curve $\gamma_S$. Therefore, MaxEnt is equivalent with MinEne for convex entropies.


\section{Finite volume discretization of binary DGM}\label{sec.fvm}
We summarize the binary, neutral, non-isothermal dusty gas model for ideal gasses described by Equations~\eqref{eq.DGM.system}, \eqref{eq.DGM2.free-energy}, and~\eqref{eq.DGM2.FF}. It is given by
\begin{subequations}
\begin{align}
    \partial_t n_\alpha + \partial_i J_\alpha^i 
    &= 0
    ~,
    \quad\mbox{for }\alpha\in\{a,b\}
    ~,
    \\    
    \partial_t \left( 
    T\left( 
        c_a^\textrm{V}n_a
       +c_b^\textrm{V}n_b
       +c_D^\textrm{V}n_D
    \right) 
    \right) 
    +
    \partial_i\left( 
        -\lambda\partial_i T
        + T\left(
            \gamma_a c_a^\textrm{V}J_a^i
          + \gamma_b c_b^\textrm{V}J_b^i
        \right)
    \right) 
    &=0
    ~.
\end{align}
where the force-flux relation are given as follows,
\begin{align} 
\begin{bmatrix}
      \partial_i n_a + n_a \partial_i \ln T
\\
    \partial_i n_b +n_b  \partial_i \ln T
\\
\end{bmatrix}
=
\underbrace{
\begin{bmatrix}
    -
    \left( 
		\frac{1}{n_a + n_b}
          \frac{n_b }{\frac{\varepsilon}{\tau}D_{ab}}
          + \frac{1}{\frac{\varepsilon}{\tau}D_{aD}}
    \right)
&
    \frac{n_a}{n_a + n_b}
    \frac{1}{\frac{\varepsilon}{\tau}D_{ab}}  \\
 \\
 \frac{n_b}{n_a+n_b}
 \frac{1}{\frac{\varepsilon}{\tau}D_{ab}}
&
    -
    \left(
        \frac{1}{\frac{\varepsilon}{\tau}D_{bD}} 
		+ \frac{1}{n_a+n_b}\frac{n_a}{\frac{\varepsilon}{\tau}D_{ab}}
    \right)
\\
\end{bmatrix}
}_{
\mathbf{M}
}
\begin{bmatrix}
J^i_{a} \\
J^i_{b} \\
\end{bmatrix}
~.
\end{align}
\label{eq:num_dgm}
\end{subequations}
Let us assume a Voronoi tessellation $\{\omega_k\}_{k=1}^N$ of the domain $\Omega$. 
For a control volume $\omega_k$, the discrete values of the densities and temperature in the collocation point $\mathbf{x_k}\in\omega_k$ are defined as 
\begin{align}
    n_{\alpha, k} 
	&:=
  \frac{1}{|\omega_k|}\int_{\omega_k} n_\alpha \mathrm{d}\mathbf{x}
	~,
	\quad\alpha\in\{a,b\}
    ~,
    \\
	T_{k}
    &:=
  \frac{1}{|\omega_k|}  \int_{\omega_k} T   \mathrm{d}\mathbf{x}
	~.
  \label{eq.discrete-system}
\end{align}
Integration over a control volume $\omega_k$ and application of the divergence theorem gives the discrete mass balance which reads
\begin{align}
|\omega_k|\partial_t n_{\alpha, k}
+\sum_{\sigma_{kl}\neq\emptyset}|\sigma_{kl}| J^{\textrm{num}}_{\alpha, kl} 
&=0
~,\quad\alpha\in\{a,b\}
\\
|\omega_k| \partial_t 
\left(
    T_{k} \left(
			c^V_a n_{a,k}
		   +c^V_b n_{b,k}
		   +c^V_D n_{D}
	\right)
\right)
+\sum_{\sigma_{kl}\neq\emptyset}|\sigma_{kl}| J^{\textrm{num}}_{T, kl}
&=
0
~,
\end{align}
where $J^{\textrm{num}}_{\alpha, kl}$ is the numerical flux of density $n_\alpha$ across the edge $\sigma_{kl} = \partial\omega_k\cap\partial\omega_k$; $J^{\textrm{num}}_{T, kl}$ is the numerical flux 
The discrete heat flux is defined as 
\begin{align}
J^{\textrm{num}}_{T, kl} 
= 
\frac{\lambda}{h_{kl}}\left[ 
	T_{l} B(Q_{T,kl})
   -T_{k} B(-Q_{T,kl})
\right]
\\
\text{where}\quad Q_{T,kl}  
= 
\frac{1}{h_{kl}\lambda}
\sum_{\alpha=a,b} \gamma_\alpha c^V_\alpha J^{\textrm{num}}_{\alpha, kl}
\nonumber
~.
\end{align}
Here $B(x) = \frac{x}{e^x-1}$ is the Bernoulli function and $h_{kl} = |\mathbf{x}_k - \mathbf{x}_l|$ denotes the distance of the collocation points.
\begin{align}
    F^{\textrm{num}}_{\alpha, kl} 
	&= 
	\frac{1}{h_{kl}}
  	\left( 
       n_{\alpha, l} B(Q_{kl})
      -n_{\alpha, k} B(-Q_{kl})
  	\right)
~,\quad\alpha\in\{a,b\}
  ~,
  \\
  \text{where}\quad  Q_{kl}  &= -\frac{1}{h_{kl}}(\log T_l - \log T_k)
~,\quad\text{for}~
k=1,\dots,N
~,
  \nonumber
~.
\end{align}
We assume the following discrete version of the mobility matrix $\mathbf{M}$
\begin{align}
M^{\textrm{num}}_{kl}
=\frac{\tau}{\varepsilon}
\begin{bmatrix}
    -
    \left( 
          \frac{\bar{n}_{b,kl} }{\bar{n}_{kl}}\frac{1}{D_{ab}}
          + \frac{1}{D_{aD}}
    \right)
&
    \frac{\bar{n}_{a,kl}}{\bar{n}_{kl}}
    \frac{1}{D_{ab}}  \\
 \\
 	\frac{\bar{n}_{b,kl}}{\bar{n}_{kl}}
 	\frac{1}{D_{ab}}
&
    -
    \left(
        \frac{\bar{n}_{a,kl}}{\bar{n}_{kl}}\frac{1}{D_{ab}}
		+\frac{1}{D_{bD}} 
    \right)
\\
\end{bmatrix}
\end{align}
where $\bar{n}_{\alpha,kl} := \tfrac{1}{2}\left(n_{\alpha,k} + n_{\alpha,l}\right)$ and $\bar{n}_{kl} := \bar{n}_{a,kl} + \bar{n}_{b,kl}$.
The linear system, 
\begin{align}
  F^{\textrm{num}}_{kl} = M^{\textrm{num}}_{kl} J^{\textrm{num}}_{kl}
  ~,
  \label{eq.implicit-fluxes}
\end{align}
gives the discrete mass fluxes.

\subsection{Numerical solution in 1D}
We wrote a \textit{Pluto.jl} notebook which shows a solution of the discrete system~\eqref{eq.discrete-system} endowed with Dirichlet boundary conditions on an interval. The solution of the discrete system itself is solved by the finite volume package~\textit{Voronoi.jl}~\cite{fuhrmann-voronoifvm}. 
Additionally, we coded a parallel solution with analytically inverted $M$, that is $J^{\textrm{num}}_{kl} = \left(M^{\textrm{num}}_{kl}\right)^{-1} F^{\textrm{num}}_{kl}$, to check the implicitly calculated fluxes in equation~\eqref{eq.implicit-fluxes}. The two approaches agree, the notebook can be downloaded here~\cite{supplement-repository}.
\begin{figure}
    \includegraphics[width=1.0\textwidth]{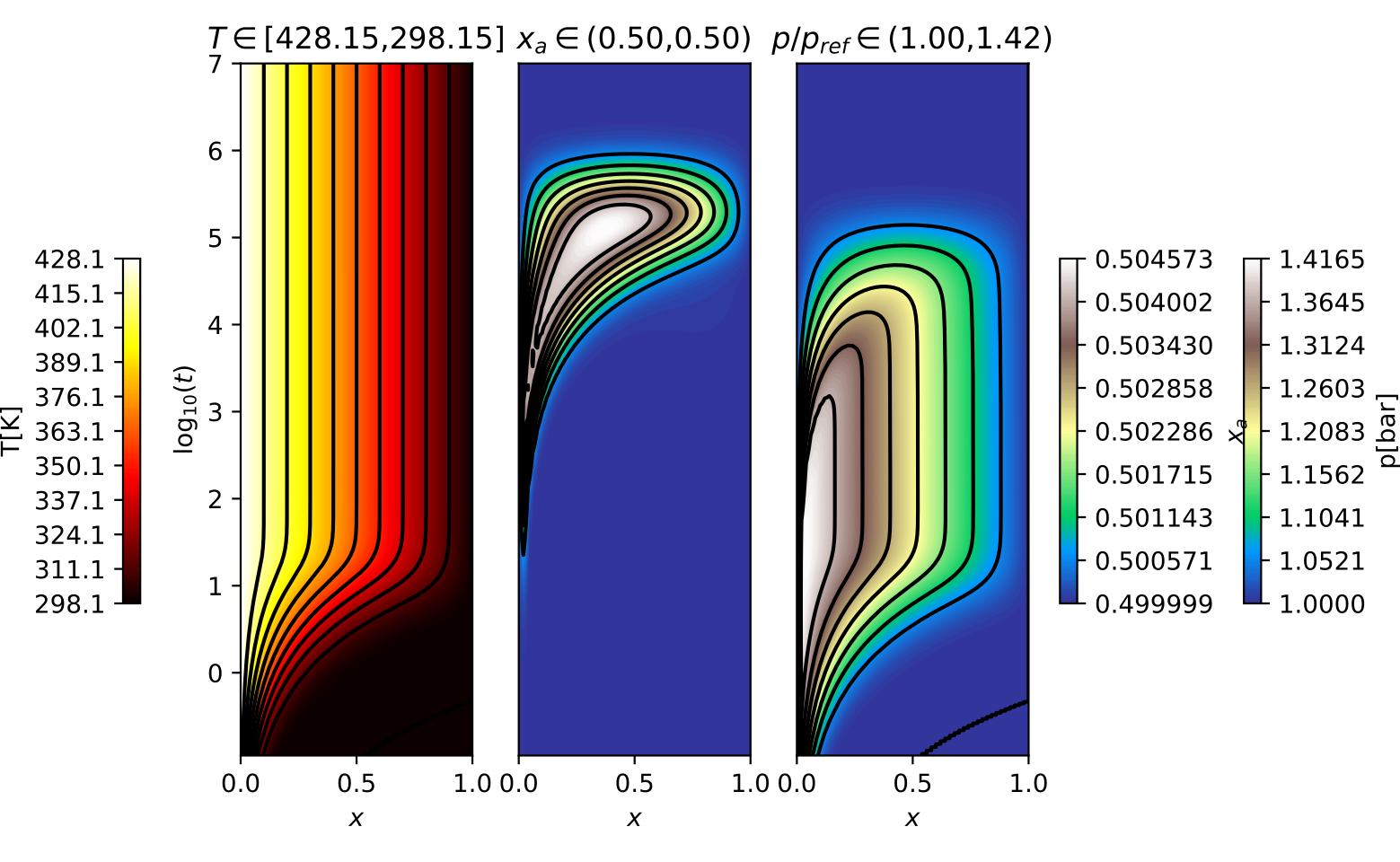}
  \label{fig:tempstep}
  \caption{Relaxation of system~\eqref{eq:num_dgm} after temperature step of the boundary condition ($\Delta T = 130\textrm{K}$). The Dirichlet boundary conditions for the pressure $p$ and the molar fraction $x_a$ are constant and equal for the both boundaries. }
\end{figure}
\end{document}